\documentclass[10pt,preprint]{aastex}
\singlespace

\shorttitle{\WMAP\ 5-year Likelihood and Parameters}
\shortauthors{Dunkley et al.}

\newcommand{\map}    {{\sl WMAP}}

\newcommand{\lt}     {\mbox{$<$}}
\newcommand{\gt}     {\mbox{$>$}} 

\newcommand{\be}{\begin{equation}}
\newcommand{\ee}{\end{equation}}
\newcommand{\ba}{\begin{eqnarray}}
\newcommand{\ea}{\end{eqnarray}}
 
\newcommand{\mb}{\mathbf}

\def\ltsima{$\; \buildrel < \over \sim \;$}
\def\ltsim{\lower.5ex\hbox{\ltsima}}
\def\gtsima{$\; \buildrel > \over \sim \;$}
\def\gtsim{\lower.5ex\hbox{\gtsima}}

\slugcomment{Astrophysical Journal Supplement Series accepted}

\begin{document}

\title{Five-Year Wilkinson Microwave Anisotropy Probe 
(WMAP\altaffilmark{1}) Observations: Likelihoods and Parameters from the WMAP data}

\author{{J. Dunkley} \altaffilmark{2,3,4}, 
{E. Komatsu} \altaffilmark{5}, 
{M. R. Nolta} \altaffilmark{6}, 
{D. N. Spergel} \altaffilmark{3,7}, 
{D. Larson} \altaffilmark{8},
{G. Hinshaw} \altaffilmark{9}, 
{L. Page} \altaffilmark{2}, 
{C. L. Bennett} \altaffilmark{8},  
{B. Gold} \altaffilmark{8}, 
{N. Jarosik} \altaffilmark{2}, 
{J. L. Weiland} \altaffilmark{10},
{M. Halpern} \altaffilmark{11}, 
{R. S. Hill} \altaffilmark{10}, 
{A. Kogut} \altaffilmark{9}, 
{M. Limon} \altaffilmark{12},
{S. S. Meyer} \altaffilmark{13},
{G. S. Tucker} \altaffilmark{14}, 
{E. Wollack} \altaffilmark{9}, 
{E. L. Wright} \altaffilmark{15}} 

\altaffiltext{1}{\map\ is the result of a partnership between Princeton 
                 University and NASA's Goddard Space Flight Center. Scientific 
		 guidance is provided by the \map\ Science Team.}

\altaffiltext{2}{{Dept. of Physics, Jadwin Hall,  Princeton University, Princeton, NJ 08544-0708}}
\altaffiltext{3}{{Dept. of Astrophysical Sciences,  Peyton Hall, Princeton University, Princeton, NJ 08544-1001}}
\altaffiltext{4}{{Astrophysics, University of Oxford,  Keble Road, Oxford, OX1 3RH, UK}}
\altaffiltext{5}{{Univ. of Texas, Austin, Dept. of Astronomy,  2511 Speedway, RLM 15.306, Austin, TX 78712}}
\altaffiltext{6}{{Canadian Institute for Theoretical Astrophysics,  60 St. George St, University of Toronto,  Toronto, ON  Canada M5S 3H8}}
\altaffiltext{7}{{Princeton Center for Theoretical Physics,  Princeton University, Princeton, NJ 08544}}
\altaffiltext{8}{{Dept. of Physics \& Astronomy,  The Johns Hopkins University, 3400 N. Charles St.,  Baltimore, MD  21218-2686}}
\altaffiltext{9}{{Code 665, NASA/Goddard Space Flight Center,  Greenbelt, MD 20771}}
\altaffiltext{10}{{Adnet Systems, Inc.,  7515 Mission Dr., Suite A100, Lanham, Maryland 20706}}
\altaffiltext{11}{{Dept. of Physics and Astronomy, University of  British Columbia, Vancouver, BC  Canada V6T 1Z1}}
\altaffiltext{12}{{Columbia Astrophysics Laboratory,  550 W. 120th St., Mail Code 5247, New York, NY  10027-6902}}
\altaffiltext{13}{{Depts. of Astrophysics and Physics, KICP and EFI,  University of Chicago, Chicago, IL 60637}}
\altaffiltext{14}{{Dept. of Physics, Brown University, 182 Hope St., Providence, RI 02912-1843}}
\altaffiltext{15}{{UCLA Physics \& Astronomy, PO Box 951547,  Los Angeles, CA 90095-1547}}

\email{j.dunkley@physics.ox.ac.uk}

\begin{abstract}
This paper focuses on cosmological constraints derived from 
analysis of \map\ data alone. 
A simple $\Lambda$CDM cosmological model fits 
the five-year \map\ temperature
and polarization data. 
The basic parameters of the model are consistent with the three-year data 
and now better constrained:
\ensuremath{\Omega_bh^2 = 0.02273\pm 0.00062}, 
\ensuremath{\Omega_ch^2 = 0.1099\pm 0.0062}, 
\ensuremath{\Omega_\Lambda = 0.742\pm 0.030}, 
\ensuremath{n_s = 0.963^{+ 0.014}_{- 0.015}}, 
\ensuremath{\tau = 0.087\pm 0.017}, and \ensuremath{\sigma_8 = 0.796\pm 0.036},
with \ensuremath{h = 0.719^{+ 0.026}_{- 0.027}}.
With five years of polarization data, we have
measured the optical depth to reionization, $\tau~\gt~0$, 
at $5\sigma$ significance. The redshift of an instantaneous 
reionization is constrained to be 
\ensuremath{z_{\rm reion} = 11.0\pm 1.4} 
with 68\% confidence. The $2\sigma$ lower limit is $z_{\rm reion}~\gt~8.2$, and 
the $3\sigma$ limit is $z_{\rm reion}~\gt~6.7$. This excludes a 
sudden reionization of the universe at $z = 6$ at more than 
$3.5\sigma$ significance, suggesting that reionization 
was an extended process.
Using two methods for polarized foreground cleaning we get consistent 
estimates for the optical depth, indicating an error due to foreground 
treatment of $\tau \sim0.01$. This cosmological model also fits small-scale CMB data, and a range of 
astronomical data measuring the expansion rate and clustering 
of matter in the universe.
We find evidence for the first time in the CMB power spectrum 
for a non-zero cosmic neutrino background, 
or a background of relativistic species, 
with the standard three light
neutrino species preferred over the best-fit $\Lambda$CDM model with 
$N_{\rm eff}=0$ at $\gt~99.5\%$ confidence, and 
\ensuremath{N_{\rm eff} > 2.3\ \mbox{(95\% CL)}} when varied. 
The five-year \map\ data improve the upper limit on the 
tensor-to-scalar ratio, \ensuremath{r < 0.43\ \mbox{(95\% CL)}}, for 
power-law models, and halve the limit on $r$ for models with a running index, 
\ensuremath{r < 0.58\ \mbox{(95\% CL)}}. With longer integration
we find no evidence for a running spectral index, with 
\ensuremath{dn_s/d\ln{k} = -0.037\pm 0.028}, and find 
improved limits on isocurvature fluctuations.
The current \map-only limit on the sum of the 
neutrino masses is \ensuremath{\sum m_\nu < 1.3\ \mbox{eV}\ \mbox{(95\% CL)}}, 
which is robust, to within 10\%, to a varying tensor amplitude, 
running spectral index or dark energy equation of state. 

\end{abstract}


\section{Introduction}
\label{sec:intro}

The {\it Wilkinson Microwave Anisotropy Probe} (\map), launched in 2001, 
has mapped out the 
Cosmic Microwave Background with unprecedented accuracy over the whole sky. 
Its observations have led to the establishment of a simple 
concordance cosmological model for the contents and evolution of 
the universe, consistent with virtually all other astronomical measurements. 
The \map\ first-year and three-year data have allowed us to 
place strong constraints on the parameters describing the
$\Lambda$CDM model, a flat universe filled with 
baryons, cold dark matter, neutrinos, and a cosmological 
constant, with initial fluctuations described by nearly 
scale-invariant power law 
fluctuations, as well as placing limits on extensions to this simple model 
\citep{spergel/etal:2003,spergel/etal:2007}.
With all-sky measurements of the polarization anisotropy
\citep{kogut/etal:2003,page/etal:2007}, two orders of magnitude smaller than the 
intensity fluctuations, \map\ has not only given us an additional picture of 
the universe as it transitioned from ionized to neutral at redshift
$z\sim1100$, but also an observation of the later reionization 
of the universe by the first stars. 

In this paper we present cosmological constraints from \map\ alone, for both the $\Lambda$CDM model and a set of possible extensions. We also 
consider the consistency of \map\ constraints with other 
recent astronomical observations. This is one of seven five-year \map\ papers. \citet{hinshaw/etal:prep} describe the data processing and basic results, 
\citet{hill/etal:prep} present new beam models and window functions, 
\citet{gold/etal:prep} describe the emission from Galactic foregrounds, and \citet{wright/etal:prep} the emission from extra-Galactic point sources. 
The angular power spectra are described in \citet{nolta/etal:prep}, and 
\citet{komatsu/etal:prep} present and interpret cosmological constraints based 
on combining \map\ with other data. 

\map\ observations are used to 
produce full-sky maps of the CMB in five frequency bands 
centered at 23, 33, 41, 61, and 94 GHz \citep{hinshaw/etal:prep}.
With five years of data, we are now able to place better limits on the 
$\Lambda$CDM model, as well as to move beyond it to test the  
composition of the universe, details of 
reionization, sub-dominant components, 
characteristics of inflation, and primordial fluctuations. 
We have more than doubled the amount of polarized data used for 
cosmological analysis, allowing a better measure of the large-scale E-mode 
signal \citep{nolta/etal:prep}. To this end we test two alternative ways 
to remove Galactic foregrounds from low resolution polarization maps, 
marginalizing over Galactic emission, providing a cross-check of our 
results. With longer integration we 
also better probe the second and third acoustic peaks in the 
temperature angular power spectrum, and have many more 
year-to-year difference maps available for cross-checking systematic effects 
\citep{hinshaw/etal:prep}.

The paper is structured as follows. 
In \S\ref{sec:like_param} we focus on the 
CMB likelihood and parameter estimation 
methodology. We describe two 
method used to clean the polarization maps, describe a fast method for 
computing the large-scale temperature 
likelihood, based on work described in
\citet{wandelt/larson/lakshminarayanan:2004}, which also uses Gibbs sampling,
and outline more efficient techniques for sampling cosmological parameters.
In \S\ref{sec:lcdm_results} we present 
cosmological parameter results from five years of 
\map\ data for the $\Lambda$CDM model, and discuss their
consistency with recent astronomical observations. Finally we consider 
constraints from \map\ alone on a set of extended cosmological models in 
\S\ref{sec:ext_results}, and conclude in \S\ref{sec:discuss}.

\section{Likelihood and parameter estimation methodology}
\label{sec:like_param}

\subsection{Likelihood}
\label{subsec:like}
The \map\ likelihood function takes the same format as for the 
three-year release, and software implementation  
is available on LAMBDA (http://lambda.gsfc.nasa.gov) 
as a standalone package. It 
takes in theoretical CMB temperature (TT), E-mode polarization (EE), 
B-mode polarization (BB), and temperature-polarization cross-correlation (TE) 
power spectra for a given cosmological model. It 
returns the sum of various likelihood components: low-$\ell$ temperature, 
low-$\ell$ TE/EE/BB polarization, high-$\ell$ temperature, high-$\ell$ TE 
cross-correlation, and 
additional terms due to uncertainty in the 
\map\ beam determination, and possible error in the extra-galactic point source removal. There is also now an additional option to compute the TB and EB 
likelihood. We describe the method used for computing the low-$\ell$ 
polarization likelihood in Section \ref{subsec:gibbs_pol}, based on 
maps cleaned by two different methods. The main improvement 
in the five-year analysis is the implementation of 
a faster Gibbs sampling method for computing the $\ell \le 32$ TT 
likelihood, which we describe in Section \ref{subsec:gibbs_tt}.

For $\ell>32$, the TT likelihood uses the combined pseudo-$C_\ell$ spectrum 
and covariance matrix described in \citet{hinshaw/etal:2007}, estimated using
V and W bands. 
We do not use the EE or BB power spectra at $\ell>23$, but continue to 
use the TE likelihood described in \citet{page/etal:2007}, estimated using
Q and V bands.
The errors due to beam and point sources are treated the same as in the 
three-year analysis, described in Appendix A of \citet{hinshaw/etal:2007}. 
A discussion of this treatment can be found in \citet{nolta/etal:prep}.

\subsubsection{Low-$\ell$ polarization likelihood}
\label{subsec:gibbs_pol}

We continue to evaluate the exact likelihood for the polarization maps at 
low multipole, $\ell \le 23$, as described in Appendix D of 
\citet{page/etal:2007}. The input 
maps and inverse covariance matrix used in the main analysis are 
produced by co-adding the template-cleaned maps described in 
\citet{gold/etal:prep}.
In both cases these are weighted to account for the P06 mask using the method 
described in \citet{page/etal:2007}.
In the three-year analysis we conservatively used only the Q and V bands in 
the likelihood. We are now confident that Ka band is cleaned sufficiently for 
inclusion in analyses (see \citet{hinshaw/etal:prep} for a discussion).

We also cross-check the polarization likelihood by using 
polarization maps obtained with an alternative component separation method, 
described in \citet{dunkley/etal:prep}.  The low resolution polarization 
maps in the 
K, Ka, Q, and V bands, degraded to HEALPix $N_{\rm side}=8$ 
\footnote{The number of pixels is $12N^2_{\rm side}$, where 
$N_{\rm side}=2^3$ for resolution 3 \citep{gorski/etal:2005}.}, are 
used to estimate a single set of marginalized Q and U CMB maps and 
associated inverse covariance matrix, that can then be used as inputs for 
the $\ell<23$ likelihood.  This is done by estimating the joint 
posterior distribution for the amplitudes and spectral indices of 
the synchrotron, dust, and CMB Q and U components, using a 
Markov Chain Monte Carlo method. One then 
marginalizes over the foreground amplitudes and spectral indices to 
estimate the CMB component. A benefit of this method is that
errors due to both instrument noise and foreground uncertainty are accounted 
for in the marginalized CMB covariance matrix. 

\subsubsection{Low-$\ell$ temperature likelihood}
\label{subsec:gibbs_tt}

For a given set of cosmological parameters with theoretical power spectrum 
$C_\ell$, the likelihood function returns $p(\mb d|C_\ell)$, the 
likelihood of the observed map $\mb d$, or its transformed $a_{lm}$ coefficients. 
Originally, the likelihood code was  written as a hybrid
combination of a normal and lognormal distribution
\citep{verde/etal:2003}.  This algorithm did not properly model the
tails of the likelihood at low multipoles
\citep{efstathiou:2004b, slosar/seljak/makarov:2004, odwyer/etal:2004,
hinshaw/etal:2007}, and so for the three-year data the $\ell \le 30$ likelihood 
was computed exactly, using
\be
p(\mb d|C_\ell)=\frac{\exp[(-(1/2)\mb d^T {\mb C}^{-1} \mb d]}{\sqrt{\det \mb C}},
\label{eqn:pixlike}
\ee
where $\mb C$ is the 
covariance matrix of the data including both the 
signal covariance matrix and noise 
$\mb C(C_\ell)=\mb S(C_\ell)+ \mb N$ 
(e.g., \citet{tegmark:1997,bond/jaffe/knox:1998,hinshaw/etal:2007}). 
This approach is computationally intensive however, since it requires 
the inversion of a large covariance matrix each time the likelihood is called.

In \citet{jewell/levin/anderson:2004,wandelt/larson/lakshminarayanan:2004,eriksen/etal:2004} a faster method was
developed and implemented to compute $p(\mb d|C_\ell)$, which we now adopt.  
It is described in
detail in those papers, so we only briefly outline the method here.  The
method uses Gibbs sampling to first sample from the joint posterior
distribution $p(C_\ell,\mb s|\mb d)$, where $C_\ell$ is the power
spectrum and $\mb s$ is the true sky signal.  From these samples, a
Blackwell-Rao (BR) estimator provides a continuous approximation to
$p(C_\ell|\mb d)$.  When a flat prior, $p(C_\ell)=const$, is used in
the sampling, we have $p(C_\ell|\mb d) \propto p(\mb d|C_\ell)$, where
the constant of proportionality is independent of $C_\ell$.  The BR
estimator can then be used as an accurate representation of the
likelihood, $p(\mb d|C_\ell)$ \citep{wandelt/larson/lakshminarayanan:2004,chu/etal:2005}.

The first step requires drawing samples from $p(C_\ell, \mb s|\mb d)$.
We cannot draw samples from the joint distribution directly, but we
can from the two conditional distributions $p(\mb s|C_\ell,\mb d)$ and
$p(C_\ell|\mb s, \mb d)$, each a slice through the $(N_{p} \times
\ell_{\rm max})$-dimensional space. Samples are drawn alternately, forming
a Markov Chain of points by the Gibbs algorithm. For the case of one 
$C_\ell$ parameter and one $s$ parameter the sampling goes as follows.
We start from some arbitrary point $(C_\ell^i,s^i)$ in the parameter space, 
and then draw 
\be
(C_\ell^{i+1}, s^{i+1}), (C_\ell^{i+2}, s^{i+2})... 
\ee
by first drawing $C_\ell^{i+1}$ from $p(C_\ell|s^i,\mb d)$ 
and then drawing $s^{i+1}$ from $p(s|C_\ell^{i+1},\mb d)$. Then 
we iterate many times. The result is a Markov chain whose stationary 
distribution is $p(C_\ell, s|\mb d)$. This is extended to vectors for 
$C_\ell$ (of length $\ell_{\rm max}$) and $\mb s$ (of length $N_p$) following the same method:
\ba 
\mb s^{i+1} &\leftarrow& p(\mb s^i|C_\ell^i,\mb d) \\ 
C_\ell^{i+1} & \leftarrow &p(C_\ell^i|\mb s^{i+1},\mb d),  
\ea
sampling the joint distribution $p(C_\ell, \mb s|\mb d)$. This sampling 
procedure is also described in \citet{jewell/levin/anderson:2004,wandelt/larson/lakshminarayanan:2004,eriksen/etal:2007}.

The first conditional distribution is a multivariate Gaussian with
mean $\mb S^i(\mb S^i+\mb N)^{-1} \mb d$ and variance $[(\mb
S^i)^{-1}+\mb N^{-1}]^{-1}$, so at each step a new signal vector $\mb
s^{i+1}$, of size $N_{p}$, can be drawn. This is computationally
demanding however, as described in
\citet{eriksen/etal:2004,wandelt/larson/lakshminarayanan:2004}, 
requiring the solution
of a linear system of equations $\mb M \mb v = \mb w$, with $\mb
M=1+\mb S^{1/2}\mb N^{-1}\mb S^{1/2}$. These are solved at each step
using the conjugate gradient technique, which is sped up by finding an
approximate inverse for $\mb M$, a preconditioner. This requires
computation of the inverse noise matrix, $\mb N^{-1}$ in spherical
harmonic space, which is done by computing the components of $\mb
N^{-1}$ term by term using spherical harmonics in pixel space.  There
are more efficient ways to compute $\mb N^{-1}$ 
\citep{hivon/etal:2002,eriksen/etal:2004e}, but computing the 
preconditioner is a one-time
expense, and it is only
necessary to compute harmonics up to $\ell = 30$.

The second conditional distribution, $p(C_\ell|\mb s,\mb d)$ is an
inverse Gamma distribution, from which a new $C_\ell$ vector of size
$\ell_{\rm max}$ can be rapidly drawn using the method in
\citet{wandelt/larson/lakshminarayanan:2004}.  Sampling from these two
conditional distributions is continued in turn until convergence, at which
point the sample accurately represents the underlying
distribution. This is checked in practice using a jacknife test that compares
likelihoods from two different samples.

Finally, once the joint distribution $p(C_\ell,\mb s|\mb d)$ has been
pre-computed, the likelihood for any given model $C_\ell$ is obtained
by marginalizing over the signal $\mb s$, $p(\mb d|C_\ell) \propto
\int p(C_\ell,\mb s|\mb d) d\mb s$, which holds for a uniform prior
distribution $p(C_\ell)$.  In practice this is computed using the
Blackwell-Rao estimator,
\be
p(\mb d|C_\ell) \propto \int  p(C_\ell|\mb s)p(\mb s|\mb d) d\mb s \approx 
\frac1{n_{\rm G}} \sum_{i=1}^{n_{\rm G}} p(C_\ell|\mb s^i)
\ee
where the sum is over all $n_{G}$ samples in the Gibbs chain.  Since
$p(C_\ell|\mb s^i) = p(C_\ell|\sigma_\ell^i)$, where
$\sigma_\ell=(2\ell+1)^{-1}\sum_m |s_{\ell m}|^2$, and $s_{\ell m}$
are the spherical harmonic coefficients of $\mb s$, one only needs to
store $\sigma_\ell^i$ at each step in the Gibbs sampling. Then, each
time the likelihood is called for a new $C_\ell$, one computes
${\cal L}=\sum_{i=1}^{n_{\rm G}} p(C_\ell|\sigma_\ell^i)/n_G$.
This requires only $O(\ell_{\rm max} n_G)$ computations, compared to the
full $O(N_{p}^3)$ evaluation of equation \ref{eqn:pixlike}.  This
speed-up also means that the exact likelihood can be used to higher
resolution than is feasible with the full evaluation, providing a more 
accurate estimation.

\begin{figure*}[t]
  \epsscale{0.6}
  \plotone{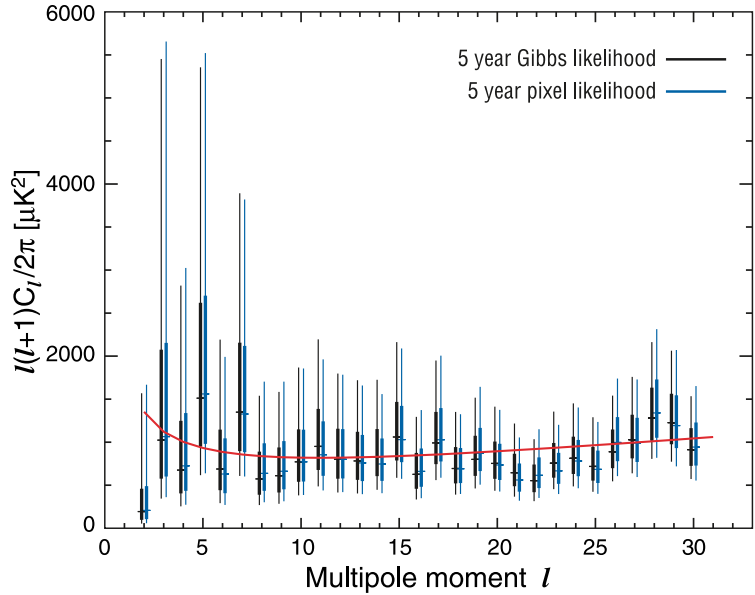}
  \caption{This figure compares the low-$\ell$ 
TT power spectrum computed with two different techniques.  
At each $\ell$ value, we plot the 
maximum likelihood value (tic mark), the region where the likelihood
is greater than 50\% of the peak value  (thick line) and 
the region where the likelihood is greater than 95\% of the 
peak value (thin line).  The black lines (left side of each pair) 
are estimated by Gibbs sampling using the ILC map smoothed with a 
5 degree Gaussian beam (at HEALPIX $N_{\rm side} = 32$).  
The light blue line (right side of the pair) is estimated with a 
pixel-based likelihood code with  $N_{\rm side} = 16$. The slight 
differences between the points are primarily due to differences in 
resolution. At each multipole, the likelihood is sampled by fixing the 
other $C_{\ell}$  values at a fiducial spectrum (red).
   \label{fig:pix_spec_check} }
\end{figure*}

\subsubsubsection{Code details: Choice of $\ell$ limits, smoothing, and 
resolution}
\label{sec:gibbsdetails}

The code used for \map\ is adapted from the MAGIC Gibbs code described 
in \citet{wandelt:2003,wandelt/larson/lakshminarayanan:2004}. The input
temperature map is the five-year ILC map described in \citet{gold/etal:prep}. 
To produce correct results, the Gibbs 
sampler requires an accurate
model of the data.  This means that the signal covariance matrix $\mb
S(C_\ell)$ cannot be approximated to be zero except for multipoles where the
smoothing makes the signal much less than the noise.
For the full \map\ data set, this would require sampling out to $\ell \sim 1000$,
with $N_{\rm side}=512$. 
This is computationally expensive, taking 
more than of order $10^4$ processor-hours to converge
\citep{odwyer/etal:2004}.  Instead we reduce the resolution and 
smooth the data to substantially reduce the required multipole 
range, speeding up the computation.  

The ILC map is smoothed to 5.0 degree FWHM, 
and sampled at $N_{\rm side}=32$. 
The process of smoothing the data has the side effect of correlating
the noise.  Properly modeling the correlated noise slows down the Gibbs 
sampling, as 
it takes longer to draw a sample from 
$p(\mb s|C_\ell,\mb d)$. We therefore
add uncorrelated white noise to the map such that it dominates 
over the smoothed noise.  However, the added
noise must not be so large that it changes the likelihood of the
low-$\ell$ modes; cosmic variance must remain dominant over the noise 
\citep{eriksen/etal:2007},  so we add 2 $\mu$K of
noise per pixel.  In Appendix A the noise power spectra are shown.

The Gibbs sampler converges much more slowly in regions of low signal
to noise.  Because of this, we only sample spectra out to $\ell=51$ and 
fix the spectrum for $51<\ell \le 96$ to a fiducial value, and set it to zero
for $\ell>96$.  
The BR estimator is only used up to $\ell=32$ for cosmological analysis, 
so we marginalize over the $32 < \ell \le 51$ spectrum.  The likelihood 
is therefore 
$p(L|d) = \int p(L,H|d) dH$, where $L$ and $H$ refer to the low, $\ell\le32$, 
and higher, $32<\ell\le 51$, parts of the power spectrum.
Examination of the BR estimator shows it to have a smooth distribution. 
Tests of the results to various input choices, including choice of resolution, 
are given in Appendix A. We 
note that by using the low-$\ell$ likelihood only 
up to $\ell \le 32$, this breaks the likelihood into a low and 
high $\ell$ part, which introduces a small error by ignoring the correlation 
between these two parts of the spectrum. However, this error is small, and 
it is unfeasible, in a realistic sampling time, to use the BR 
estimator over the entire $\ell$ range probed by \map.

In Figure \ref{fig:pix_spec_check} we show slices through 
the $C_{\ell}$ distribution 
obtained from the BR estimator, compared to the pixel likelihood code. 
The estimated spectra agree well. Some small discrepancies are due to 
the pixel code using $N_{\rm side}=16$, compared to the higher resolution 
$N_{\rm side}=32$ used for the Gibbs code. 

\subsection{Parameter Estimation}
\label{sec:param}

\begin{table}[t] 
\linespread{1.25}
\begin{center}{\footnotesize\begin{tabular}{cl}
\hline
\hline
Parameter & Description \\
\hline
$\omega_{b}$ &Baryon density,  $\Omega_{b}h^{2}$ \\
$\omega_{c}$ &Cold dark matter density,  $\Omega_{c}h^{2}$  \\
$\Omega_{\Lambda}$ & Dark energy density, with $w=-1$ unless stated \\
$\Delta_{\mathcal R}^{2}$ & Amplitude of curvature perturbations at $k_0=0.002/$Mpc\\
$n_{s}$ &Scalar spectral index at $k_0=0.002/$Mpc  \\
$\tau$  & Reionization optical depth  \\
$A_{SZ}$ & SZ marginalization factor \\
\hline
$d n_{s} /d \ln k$ & Running in scalar spectral index \\
$r$ & Ratio of the amplitude of tensor fluctuations to scalar fluctuations\\
$\alpha_{-1}$ & Fraction of anti-correlated CDM isocurvature (See Sec \ref{subsubsec:iso})\\
$\alpha_0$ & Fraction of uncorrelated CDM isocurvature (See Sec \ref{subsubsec:iso})\\
$N_{\rm eff}$ & Effective number of relativistic species (assumed neutrinos) \\
$\omega_{\nu}$ & Massive neutrino density, $\Omega_\nu h^2$ \\
$\Omega_{k}$ & Spatial curvature, $1-\Omega_{\rm tot}$ \\ 
$w$ & Dark energy equation of state, $w= p_{DE}/\rho_{DE}$\\
$Y_P$ & Primordial Helium fraction\\
$x_e$ & Ionization fraction of first step in two-step reionization \\
$z_r$ & Reionization redshift of first step in two-step reionization \\
\hline
$\sigma_8$ & Linear theory amplitude of matter fluctuations on 8 $h^{-1}$ Mpc scales \\	
$H_{0}$ & Hubble expansion factor ($100h$ Mpc$^{-1}$km\ s$^{-1}$) \\
$\sum m_{\nu}$ & Total neutrino mass (eV) $\sum m_{\nu}=94 \Omega_{\nu}h^2$ \\
$\Omega_{m}$ & Matter energy density $\Omega_{b}+\Omega_{c}+\Omega_{\nu}$\\ 
$\Omega_{m} h^2$ & Matter energy density \\ 
$t_0$ & Age of the universe (billions of years)\\
$z_{\rm reion}$ & Redshift of instantaneous reionization\\
$\eta_{10}$ & Ratio of baryon to photon number densities, $10^{10}(n_b/n_\gamma) = 273.9~\Omega_b h^2$\\  
\hline
\end{tabular}}\end{center}
\caption{\small{Cosmological parameters used in the analysis. http://lambda.gsfc.nasa.gov lists the marginalized values for these parameters for all of the models discussed in this paper.} 
\label{table:param_def}}
\linespread{1}
\end{table}

We use the Markov Chain Monte Carlo methodology described in 
\citet{spergel/etal:2003,spergel/etal:2007} to 
explore the probability distributions for various cosmological models, 
using the five-year \map\ data and other cosmological data sets. We map 
out the full distribution for each cosmological 
model, for a given data set or data combination, and quote
parameter results using the means and 68\% confidence limits of the 
marginalized distributions, with
\be
\langle x_i \rangle=\int d^N x L(d|x) p(x) x_i = \frac{1}{M}\sum_{j=1}^{M}x_i^j
\ee
where $x_i^j$ is the $j$th value of the $i$th parameter in the chain. We also
give 95\% upper or lower limits when the distribution is one-tailed. 
We have made a number of changes in the five-year 
analysis, outlined here and in Appendix B.

We parameterize our basic $\Lambda$CDM 
cosmological model in terms of the following parameters:
\be
\{\omega_b, \omega_c, \Omega_\Lambda, \tau, n_s, \Delta_{\cal R}^2\}, \{A_{SZ}\}
\ee
defined in Table \ref{table:param_def}. 
$\Delta_{\cal R}^2$ is the amplitude of curvature perturbations, 
and $n_s$ the spectral tilt, both defined at pivot scale $k_0=0.002/$Mpc. In 
this simplest model we assume `instantaneous' reionization of the universe, 
with optical depth $\tau$, in which the 
universe transitions from being neutral to fully ionized during a change in 
redshift of $z=0.5$. 
The contents of the Universe, assuming a flat geometry, 
are quantified by the 
baryon density $\omega_b$, the cold dark matter (CDM) density $\omega_c$ 
and a cosmological constant, $\Omega_\Lambda$. We treat the 
contribution to the power 
spectrum by Sunyaev-Zeldovich fluctuations \citep{sunyaev/zeldovich:1970}
as in \citet{spergel/etal:2007}, adding the predicted template spectrum 
from \citet{komatsu/seljak:2002}, multiplied by an amplitude $A_{SZ}$, 
to the total spectrum. This template spectrum is scaled with 
frequency according to the known SZ frequency dependence. 
We limit $0<A_{SZ}<2$ and  impose unbounded 
uniform priors on the remaining six parameters.

We also consider extensions to this model, parameterized by
\be
\{dn_s/d\ln k,r, \alpha_{-1}, \alpha_{0},\Omega_K, w, \omega_\nu, N_{\rm eff}, Y_P, 
x_e, z_r\}
\ee
also defined in Table \ref{table:param_def}. These include cosmologies in 
which the primordial perturbations have a running scalar spectral index 
$dn_s/d \ln k$, a tensor contribution with 
tensor-to-scalar ratio $r$, or an anti-correlated or uncorrelated 
isocurvature component, quantified by $\alpha_{-1}$, $\alpha_0$. They also 
include models with a curved geometry $\Omega_k$, a constant 
dark energy equation of state $w$, and those with massive neutrinos 
$\sum m_\nu=94\Omega_\nu h^2~\rm eV$, 
varying numbers of relativistic species $N_{\rm eff}$, and varying 
primordial Helium fraction $Y_P$. There are also models with 
non-instantaneous `two-step' reionization as in \citet{spergel/etal:2007}, 
with an initial ionized step at $z_r$ with ionized fraction $x_e$, followed 
by a second step at $z=7$ to a fully ionized universe. 

These parameters all take uniform priors, and 
are all sampled directly, but
we bound $N_{\rm eff}<10$, $w>-2.5$, $z_r < 30$ and impose 
positivity priors on $r$, $\alpha_{-1}$, $\alpha_0$, $\omega_\nu$, $Y_p$, and 
$\Omega_\Lambda$, as well as requiring
$0<x_e<1$. The tensor spectral index is fixed at $n_t=-r/8$. 
We place a prior on the 
Hubble constant of $20<H_0<100$, but this only affects non-flat models. 
Other parameters, including $\sigma_8$, the redshift of reionization, 
$z_{\rm reion}$, and the age of the universe, $t_0$, are derived from these primary parameters and described in Table
\ref{table:param_def}. A more extensive set of derived parameters are provided on 
the LAMBDA website. In this paper we assume that the primordial 
fluctuations are Gaussian, and do not consider constraints on 
parameters describing deviations from Gaussianity in the data. 
These are discussed in the cosmological interpretation presented in
\citep{komatsu/etal:prep}.

We continue to use the CAMB code \citep{lewis/challinor/lasenby:2000}, based 
on the CMBFAST  code \citep{seljak/zaldarriaga:1996}, 
to generate the CMB power spectra for a given set of cosmological parameters
\footnote {We use the pre-March 2008 version of CAMB that treats reionization as `instantaneous'  (with width $z\sim0.5$) and fully ionizes hydrogen but not helium at reionization.}.
Given the improvement in the \map\ data, 
we have determined that distortions to 
the spectra due to weak gravitational lensing  
should now be included (see e.g. \citet{blanchard/schneider:1987,seljak:1997,hu/okamoto:2002}. 
We use the lensing option in CAMB which roughly doubles the time 
taken to generate a model, compared to the unlensed case. 

\begin{figure*}[t]
  \epsscale{0.7}
  \plotone{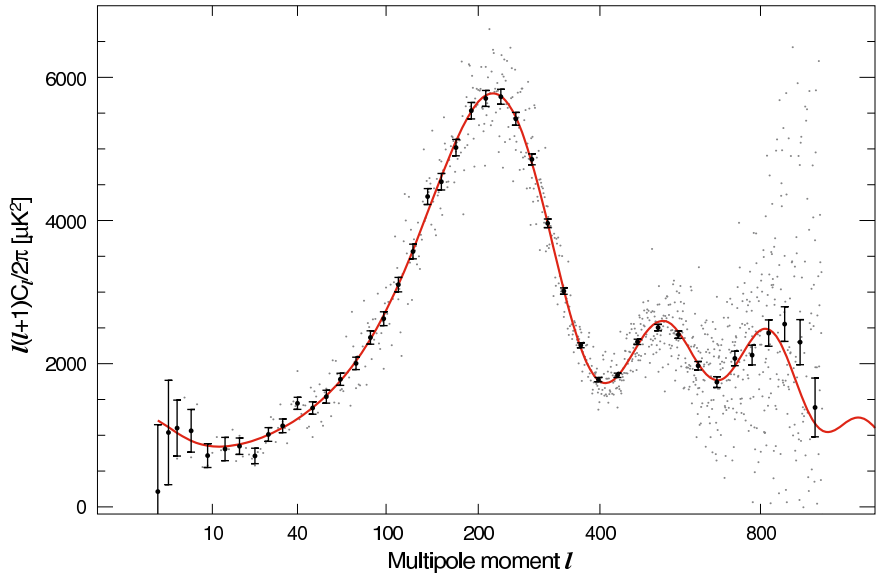}
  \caption{The temperature angular power spectrum corresponding to the \map-only 
best-fit $\Lambda$CDM model. The grey dots are the unbinned data; the black data 
points are binned data with 1$\sigma$ error bars including both noise and 
cosmic variance computed for the best-fit model. 
  \label{fig:lcdm_spec} }
\end{figure*}

We have made a number of changes in the parameter sampling methodology. Our 
main pipeline now uses an MCMC code originally developed for use in 
\citet{bucher/etal:2004}, which has been adapted for \map. 
For increased speed and reliability, it incorporates two  
changes in the methodology described in \citet{spergel/etal:2007}.
It uses a modified sampling method that generates a single chain for each 
model (instead of the four, or eight, commonly used in cosmological analyses). 
We also use an alternative spectral convergence test that 
can be run on a single chain, developed in \citet{dunkley/etal:2005}, 
instead of the Gelman \& Rubin test used in \citet{spergel/etal:2007}. 
These are both described in Appendix B. 
We also use the publicly available CosmoMC 
sampling code \citep{lewis/bridle:2002} as a secondary pipeline, used as
an independent cross-check for a limited set of models.

\section{The $\Lambda$CDM Cosmological Model}
\label{sec:lcdm_results}
\subsection{\map\ five-year parameters}

The $\Lambda$CDM model, described by just six parameters, is 
still an excellent fit to the \map\ data. The temperature and polarization 
angular power spectra are shown in 
\citet{nolta/etal:prep}. With more observation the errors on 
the third acoustic peak in the temperature angular 
power spectrum have been reduced.  
The TE cross-correlation spectrum has also 
improved, with a better
measurement of the second anti-correlation at $\ell \sim 500$. 
The low-$\ell$ signal in the EE spectrum, due to 
reionization of the universe \citep{ng/ng:1995,zaldarriaga/seljak:1997}, 
is now measured with higher significance 
\citep{nolta/etal:prep}. 
The best-fit 6 parameter model, shown in Figure \ref{fig:lcdm_spec}, 
is successful in fitting three TT acoustic 
peaks, three TE cross-correlation maxima/minima, and 
the low-$\ell$ EE signal. The model is compared to the polarization data in 
\citet{nolta/etal:prep}.
The consistency of both the temperature and polarization signals with 
$\Lambda$CDM 
continues to validate the model. 

\begin{table} [t] 
\begin{center}
\begin{tabular}{cccc}
\hline
\hline
Parameter & 3 Year Mean & 5 Year Mean & 5 Year Max Like  \\
\hline
$100\Omega_b h^2$ & $2.229\pm0.073$ & \ensuremath{2.273\pm 0.062} & 2.27\\
$\Omega_c h^2$ & $0.1054\pm0.0078$& \ensuremath{0.1099\pm 0.0062}& 0.108\\
$\Omega_\Lambda$ & $0.759\pm0.034$ & \ensuremath{0.742\pm 0.030}& 0.751\\
$n_s$ & $0.958\pm0.016$ &\ensuremath{0.963^{+ 0.014}_{- 0.015}}& 0.961\\
$\tau$ & $0.089\pm 0.030$ &\ensuremath{0.087\pm 0.017}& 0.089\\ 
$\Delta_{\cal R}^2$ & $(2.35 \pm0.13) \times 10^{-9}$ &\ensuremath{(2.41\pm 0.11)\times 10^{-9}} & 2.41  $\times 10^{-9}$\\ 
\hline
$\sigma_8$ & $0.761\pm0.049$ &\ensuremath{0.796\pm 0.036} & 0.787\\
$\Omega_m$ & $0.241\pm0.034$ &\ensuremath{0.258\pm 0.030}& 0.249\\
 $\Omega_m h^2$ & $0.128\pm0.008$ &\ensuremath{0.1326\pm 0.0063} & 0.131\\
$H_0$ & $73.2^{+3.1}_{-3.2}$ &\ensuremath{71.9^{+ 2.6}_{- 2.7}} & 72.4\\
$z_{\rm reion}$ & $11.0\pm2.6$ &\ensuremath{11.0\pm 1.4} & 11.2\\
$t_0$ & $ 13.73\pm0.16$ &\ensuremath{13.69\pm 0.13} & 13.7\\
\hline
\hline
\end{tabular}
\caption{\small{$\Lambda$CDM model parameters and 68\% confidence intervals from the five-year \map\ data alone.  The three-year values are shown for comparison.
For best estimates of parameters, the marginalized `Mean' values 
should be used. The `Max Like' values correspond to the single model 
giving the highest likelihood.}
\label{table:lcdm_params}}
\end{center}
\end{table}

\begin{figure*}[t]
  \epsscale{0.9}
  \plotone{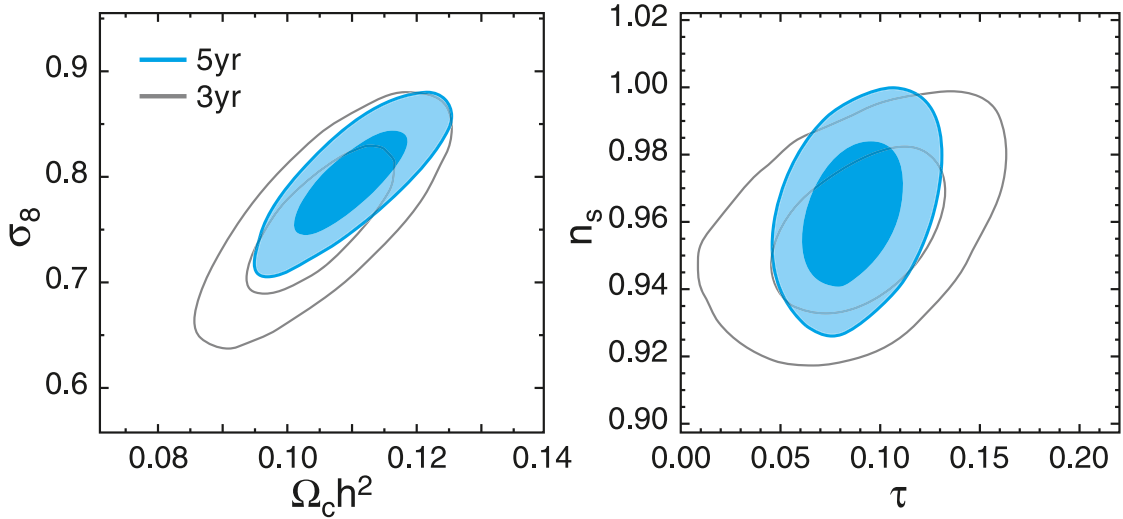}
  \caption{Constraints on $\Lambda$CDM parameters from the five-year \map\ 
data. The two-dimensional 68\% and 95\% marginalized limits are shown in blue. 
They are consistent with the three-year constraints (grey).  
Tighter limits on the amplitude of matter fluctuations, $\sigma_8$, and 
the cold dark matter 
density $\Omega_c h^2$, arise from a better measurement of 
the third temperature (TT) acoustic peak. The improved measurement of the
EE spectrum provides a 5$\sigma$ detection of the optical depth to reionization, $\tau$, which is now almost
uncorrelated with the spectral index $n_s$.
  \label{fig:twod_lcdm} }
\end{figure*}

\begin{figure*}[t]
  \epsscale{1.0}
  \plotone{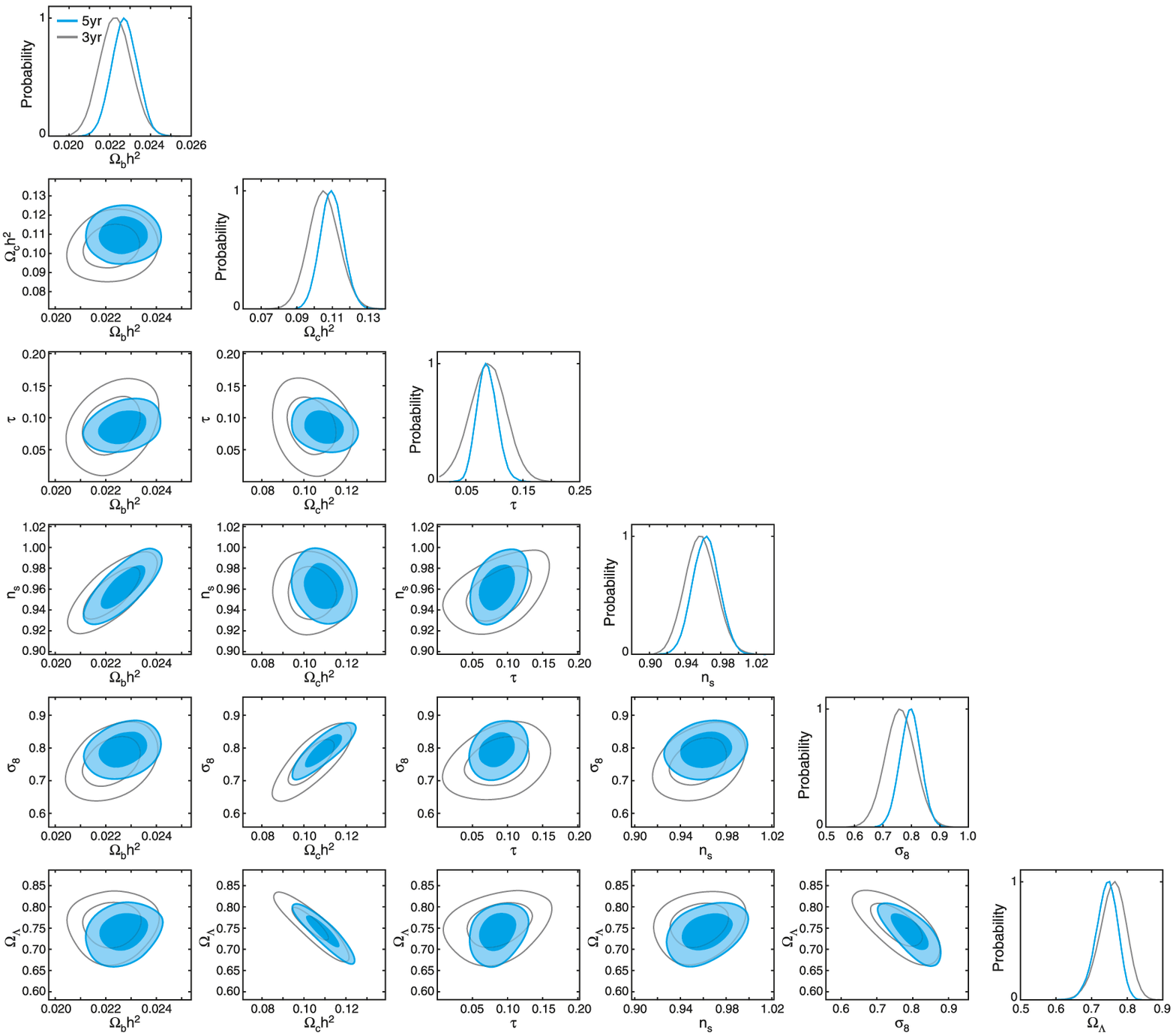}
  \caption{Constraints from the five-year \map\ data on $\Lambda$CDM 
parameters (blue), showing marginalized one-dimensional distributions and
two-dimensional 68\% and 95\% limits. 
Parameters are consistent with the three-year limits (grey) from 
\citet{spergel/etal:2007}, and are now better constrained. 
  \label{fig:oned_lcdm} }
\end{figure*}

The five-year marginalized distributions for $\Lambda$CDM, 
shown in Table \ref{table:lcdm_params} and Figures 
\ref{fig:twod_lcdm} and \ref{fig:oned_lcdm}, are consistent with the 
three-year results \citep{spergel/etal:2007}, but the uncertainties are all
reduced, significantly so for certain parameters.
With longer integration of the large-scale polarization anisotropy, 
there has been a significant improvement in the measurement of the 
optical depth to reionization. There is now a 5$\sigma$ 
detection of $\tau$, with mean value 
\ensuremath{\tau = 0.087\pm 0.017}. This can be compared to 
the three-year measure of $\tau=0.089\pm0.03$. The central value is 
little altered with two more years of integration, and the 
inclusion of the 
Ka band data, but the limits have almost halved. 
This measurement, and its implications, are discussed in 
Sec \ref{subsec:res_reion}.

The higher acoustic peaks in the TT and TE power spectra also provide more 
information about the $\Lambda$CDM model. 
Longer integration has resulted in a better measure of the 
height and position of the third peak. The highest multipoles have
a slightly higher mean value relative to the 
first peak, compared to the three-year data. This can be attributed partly 
to improved beam modeling, and partly to longer integration time reducing the noise.
The third peak position constrains 
$\Omega_m^{0.275}h$, while the third peak height strongly constrains the matter density, $\Omega_m h^2$ \citep{hu/white:1996,hu/sugiyama:1995}.  In this region of the spectrum, the WMAP data are noise-dominated so that the errors on the angular power spectrum shrink as 1/t.  The uncertainty on the matter density has dropped from 12\% in the first year data to 8\% in the three year data and now 6\% in the five year data.
The CDM density constraints are compared to three-year limits 
in Figure \ref{fig:twod_lcdm}. 
The spectral index still has a mean value 2.5$\sigma$ less than unity, with 
\ensuremath{n_s = 0.963^{+ 0.014}_{- 0.015}}. This continues to 
indicate the preference of a red spectrum consistent with the 
simplest inflationary scenarios \citep{linde:2005,boyle/steinhardt/turok:2006}, and our confidence will
be enhanced with more integration time.

Both the large scale EE spectrum and the small scale TT spectrum contribute to
an improved measure of the amplitude of matter fluctuations. With the CMB  
we measure the amplitude of curvature fluctuations, quantified by
$\Delta_{\cal R}^2$, but we also derive limits on $\sigma_8$, 
the amplitude of matter fluctuations on $8 h^{-1} \rm Mpc$ scales.
A higher value for $\tau$ produces more overall damping of the CMB 
temperature signal, making it somewhat degenerate with the amplitude, $\Delta_{\cal R}^2$, and therefore $\sigma_8$.
The value of $\sigma_8$ also affects the height of the acoustic peaks at small 
scales, so information is gained from both temperature and polarization.
The five-year data give \ensuremath{\sigma_8 = 0.796\pm 0.036}, slightly 
higher than the three-year result, driven by the increase in the amplitude of 
the power spectrum near the 
third peak. The
value is now remarkably 
consistent with new measurements from weak lensing surveys, as 
discussed in Section \ref{subsec:data}.

\subsubsection{Reionization}
\label{subsec:res_reion}

Our observations of the acoustic peaks in the TT and  TE  spectrum  
imply that most of the ions and electrons in the universe combined to  
make neutral hydrogen and helium at $z \simeq 1100$.  Observations of  
quasar spectra show diminishing Gunn-Peterson troughs at $z < 5.8$ \citep{fan/etal:2000,fan/etal:2001} implying that the universe was nearly fully ionized by $z  
= 5.7$.
How did the universe make the transition from being nearly fully  
neutral to fully ionized? The astrophysics of reionization 
has been a very active area of  
research in the past decade.  Several recent reviews \citep{barkana/loeb:2006,fan/carilli/keating:2006,furlanetto/oh/briggs:2006,meiksin:2007} summarize  
the current observations and theoretical models.  Here, we highlight a  
few of the important issues and discuss some of the implications of the \map\  
measurements of optical depth.

What objects reionized the universe?  While high redshift galaxies are  
usually considered the 
most likely source of reionization, AGNs may also have played an  
important role.
As galaxy surveys push towards ever higher redshift, it is unclear  
whether the known
population of star forming galaxies at $z \sim 6$ could have ionized  
the universe
(see e.g., \citet{bunker/etal:2007}).  The EE
signal clearly seen in the \map\ five-year data (\citet{nolta/etal:prep}; \S 2)  
implies
an optical depth, $\tau\simeq 0.09$.   This large optical depth  
suggests that higher
redshift galaxies, perhaps the low luminosity sources appearing in  $z  
 > 7$ surveys
\citep{stark/etal:2007}, played an important role in reionization.
While the known
population of AGNs can not be a significant source of reionization  
\citep{bolton/haehnelt:2007,srbinovsky/wyithe:2007}, 
an early generation of supermassive black  
holes could have played a role in reionization 
\citep{ricotti/ostriker:2004,ricotti/ostriker/mack:2007}.  
This early reionization would also have an impact on the CMB.

\clearpage

\begin{figure*}[t]
  \epsscale{1.0}
  \plotone{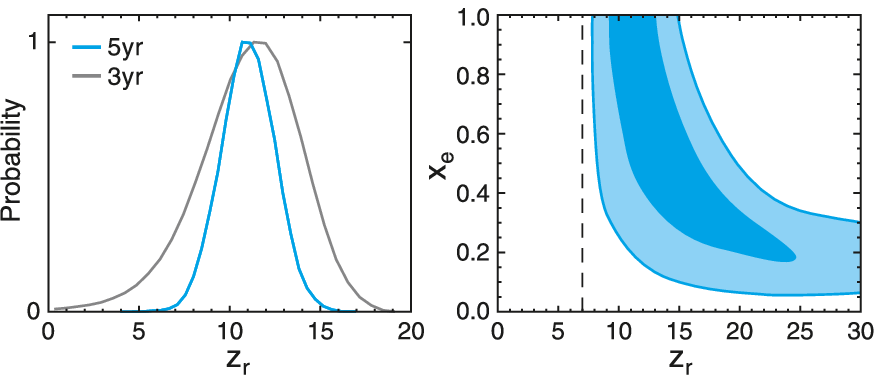}
  \caption{Left: Marginalized probability distribution for 
$z_{\rm reion}$ in the standard model with instantaneous reionization. 
Sudden reionization at $z=6$ is ruled out at 3.5$\sigma$, suggesting that
reionization was a gradual process. Right: In a model with 
two steps of reionization (with ionization fraction $x_e$ at redshift $z_r$, followed by full ionization at $z=7$), the \map\ data are consistent 
with an extended reionization process. 
  \label{fig:twod_xe} }
\end{figure*}

Most of our observational constraints probe the end of the epoch of  
reionization. Observations of  $z > 6$ quasars
\citep{becker/etal:2001,djorgovski/etal:2001,fan/carilli/keating:2006,willott/etal:2007} find that the Lyman-$\alpha$ optical  
depth rises rapidly.  Measurements of the afterglow spectrum of a  
gamma ray burst at z = 6.3 \citep{totani/etal:2006}
suggest that universe was mostly ionized at $z = 6.3$.
Lyman alpha emitter surveys \citep{taniguchi/etal:2005,malhotra/rhoads:2006, kashikawa/etal:2006,iye/etal:2006,ota/etal:prep} 
imply a significant ionized fraction at $z = 6.5$.
The interpretation that there is a sudden change in the
properties of the IGM remains a subject of active debate 
\citep{becker/rauch/sargent:2007,wyithe/bolton/haehnelt:2008}.

The \map\ data place new constraints on the reionization history of the  
universe.  
The \map\ data  best constrains the optical depth due to 
reionization at moderate  
redshift ($z < 25$) and only indirectly constrains the redshift of  
reionization.  If reionization is sudden, then the \map\ data implies  
that \ensuremath{z_{\rm reion} = 11.0\pm 1.4}, shown in 
Figure \ref{fig:twod_xe},
and now excludes $z_{\rm reion} = 6$ at more than 99.9\% CL.
The combination of the \map\ data implying that the
universe was mostly reionized at $z \sim 11$ and the measurements of  
rapidly rising optical
depth at $z \sim 6-6.5$ suggest that reionization was an extended  
process rather than
a sudden transition.  Many early studies of reionization envisioned a  
rapid transition from a neutral to a fully ionized universe occurring  
as ionized bubbles percolate and overlap.  As Figure \ref{fig:twod_xe}
shows, the \map\ data suggests a more gradual process with reionization  
beginning perhaps as early as $z \sim 20$ and strongly 
favoring $z > 6$. {\it  This suggests that the universe
underwent an extended period of  partial reionization.} 
The limits were found by modifying the
ionization history in CAMB to include two steps in the ionization 
fraction at late times ($z < 30$): the first at $z_r$ with ionization fraction 
$x_e$, the second at $z=7$ with $x_e=1$.
Several studies \citep{cen:2003,chiu/fan/ostriker:2003,wyithe/loeb:2003,haiman/holder:2003,yoshida/bromm/hernquist:2004,choudhury/ferrara:2006,iliev/etal:2007,wyithe/bolton/haehnelt:2008} suggest that feedback  
produces a prolonged or perhaps even, multi-epoch reionization history.

While the current \map\ data constrain the optical depth of the  
universe, the EE data does not yet
provide a detailed constraint on the reionization history.  With more data  
from \map\ and upcoming data
from Planck, the EE spectrum will begin to place stronger constraints  
on the details
of reionization \citep{kaplinghat/etal:2003,holder/etal:2003,mortonson/hu:2008}.  These measurements will be supplemented by measurements  
of the Ostriker-Vishniac effect by
high resolution CMB experiments which is sensitive to $\int n_e^2 dt$  
\citep{ostriker/vishniac:1986,jaffe/kamionkowski:1998, gruzinov/hu:1998}, and discussed in e.g., 
\citet{zhang/pen/trac:2004}.

\begin{figure*}[t]
  \epsscale{1.0}
  \plotone{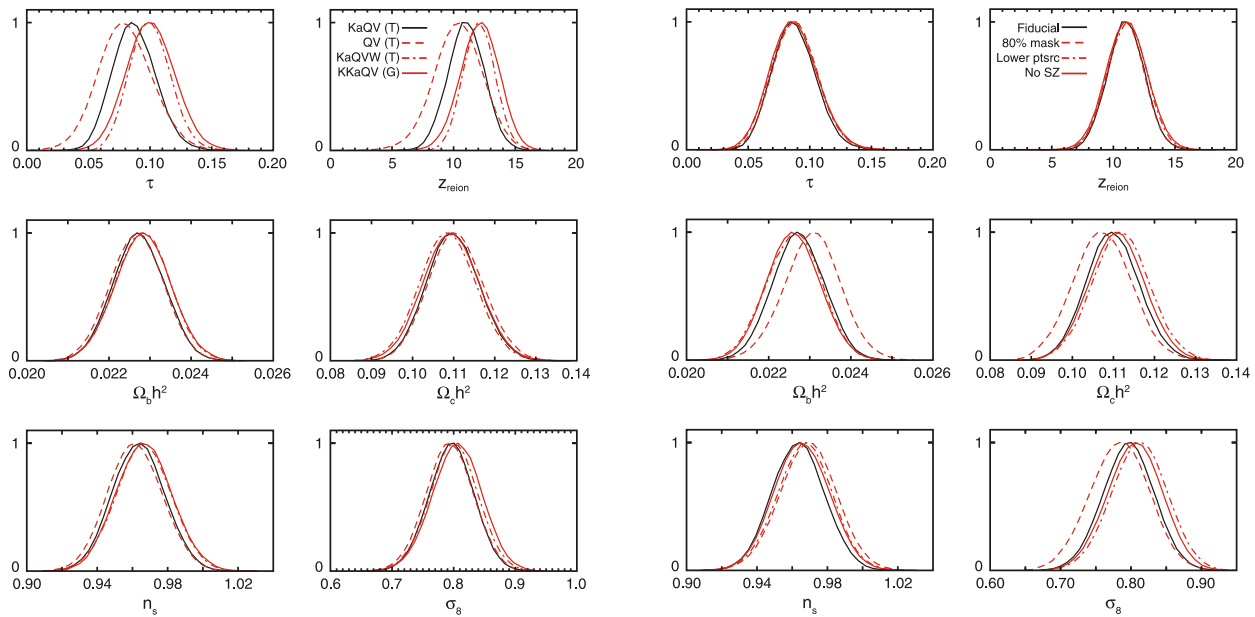}
  \caption{Effect of foreground treatment and likelihood 
details on $\Lambda$CDM parameters. Left: The number of bands used in 
the template cleaning (denoted `T'), affects the precision to which $\tau$ is 
determined, with the standard KaQV compared to QV and KaQVW, but 
has little effect on other cosmological parameters. Using maps cleaned by Gibbs
sampling (KKaQV (G)) also gives consistent results.
Right: Lowering the residual point source contribution (`lower ptsrc') and 
removing the marginalization over an SZ contribution (`No SZ') affects 
parameters by $\lt~0.4\sigma$. Using a larger mask (`80\% mask') has
a greater effect, increasing $\Omega_b h^2$ by $0.5\sigma$, but is 
consistent with the effects of noise.
\label{fig:lcdm_pol} }
\end{figure*}

\subsubsection{Sensitivity to foreground cleaning}
\label{subsec:lcdm_foreg}

As the E-mode signal is probed with higher accuracy, it becomes increasingly
important to test how much the constraint on $\tau$, $z_{\rm reion}$, and 
the other cosmological parameters, depend on details of the 
Galactic foreground 
removal. Tests were done in \citet{page/etal:2007} to show that $\tau$
was insensitive to a set of variations in the dust template used to clean the 
maps. In Figure \ref{fig:lcdm_pol} we show the 
effect on $\Lambda$CDM parameters of changing the number of bands 
used in the template-cleaning method: discarding Ka band in the `QV' 
combination, or adding W band in the `KaQVW' combination.
We find that $\tau$ (and therefore $z_{\rm reion}$)
is sensitive to the maps, but the dispersion is consistent with noise. 
As expected, the error bars are broadened for the QV combined data, 
and the mean value is $\tau=0.080\pm0.020$. 
When W band is included, the mean value is  $\tau=0.100\pm0.015$. 
We choose not to use the W band map in our main analysis however, 
because there appears 
to be excess power in the cleaned map at $\ell=7$. This indicates a potential 
systematic error, and is discussed further in \citet{hinshaw/etal:prep}.
The other cosmological 
parameters are only mildly sensitive to the number of bands used. 
This highlights the fact that $\tau$ is no longer as strongly 
correlated with other parameters, as in earlier \map\ data 
\citep{spergel/etal:2003,spergel/etal:2007}, 
notably with the spectral index of primordial 
fluctuations, $n_s$ (Figure \ref{fig:twod_lcdm}). 
 
We also test the parameters obtained using the 
Gibbs-cleaned maps described in \citet{dunkley/etal:prep}
and Section \ref{subsec:gibbs_pol}, which use the K, Ka, Q, and V band maps. 
Their distributions are 
also shown in Figure \ref{fig:lcdm_pol}, and 
have mean $\tau=0.100\pm0.018$. This is 
less than 1$\sigma$ higher than the KaQV template-cleaned maps but uses an 
independent method. The other cosmological 
parameters are changed by less than $0.3\sigma$ compared to the 
template-cleaned results. This consistency gives us confidence that
the parameter constraints are little affected by foreground uncertainty. The 
difference in central values from the two methods indicates an error
due to foreground removal uncertainty on $\tau$, in addition to the 
statistical error, of $\sim0.01$.

\subsubsection{Sensitivity to likelihood details}
\label{subsec:lcdm_like}

The likelihood code used for cosmological analysis has a number of variable
components that have been fixed using our best estimates. Here we consider 
the effect of these choices on the five-year $\Lambda$CDM parameters. The first two 
are the treatment of the residual point sources, and the treatment of the beam 
error, both discussed in \citet{nolta/etal:prep}. 
The multi-frequency data are used to estimate a residual point 
source amplitude of 
$A_{ps}=0.011 \pm 0.001~\mu \rm K^2 \rm sr$, which scales the 
expected contribution to the cross-power spectra of sources below 
our detection threshold. It is defined in 
\citet{hinshaw/etal:2007,nolta/etal:prep}, and is marginalized 
over in the likelihood code. The 
estimate comes from QVW data, whereas the VW data 
give $0.007\pm0.003~\mu \rm K^2 \rm sr$, both using the KQ85 
mask described in \citet{hinshaw/etal:prep}. 
The right panels in Figure \ref{fig:lcdm_pol} shows the effect on a subset of 
parameters of lowering $A_{ps}$ to the VW value, which leads to a slightly 
higher $n_s$, $\Omega_ch^2$ and $\sigma_8$, all within 
$0.4\sigma$ of the fiducial 
values, and
consistent with more of the observed high-$\ell$ signal being due to 
CMB rather than unresolved point sources. 
We also use $A_{ps}=0.011~\mu \rm K^2 \rm sr$ with no point source error, and find a
negligible effect on parameters ($< 0.1\sigma$). The beam window function error is 
quantified by ten modes, and in the standard treatment we 
marginalize over them, following the prescription in \citet{hinshaw/etal:2007}. We
find that removing the beam error also has a negligible effect on 
parameters. This is discussed further in \citet{nolta/etal:prep}, who 
considers alternative treatments of the beam and point source errors.

The next issue is the treatment of a possible contribution from 
Sunyaev Zeldovich fluctuations. We account for the SZ effect in the same 
way as in the three-year analysis, marginalizing over the amplitude of 
the contribution parameterized by the Komatsu-Seljak 
model \citep{komatsu/seljak:2002}. 
The parameter $A_{SZ}$ is unconstrained by the \map\ data, but is not
strongly degenerate with any other parameters.
In Figure \ref{fig:lcdm_pol} we show the effect on parameters of
setting the SZ contribution to zero. Similar to the effect of 
changing the point source contribution, the parameters depending 
on the third peak are slightly affected, with a $< 0.25\sigma$ increase 
in $n_s$, $\Omega_ch^2$, $\sigma_8$ and similar decrease in baryon density.

Another choice is the area of sky used for cosmological 
interpretation, or how much we mask out to account for Galactic contamination. 
\citet{gold/etal:prep} discuss the new masks used for the five-year analysis, 
with the KQ85 mask used as standard. We test the effect of using the 
more conservative KQ80 mask, and find a more noticeable shift. 
The quantity $\Omega_bh^2$ is increased by 0.5$\sigma$, and $n_s$, $\Omega_ch^2$ and $\sigma_8$ 
all decreased by $\sim0.4\sigma$.
This raised concerns that the KQ85 mask contains residual 
foreground contamination, but 
as discussed in \citet{nolta/etal:prep}, this shift is 
found to be consistent with the effects of noise, tested with simulations. 
We also confirm that the effect on parameters is even less for 
$\Lambda$CDM models using \map\ with external 
data, and that the choice of mask has only a small effect on the 
tensor amplitude, raising the 95\% confidence level by $\sim$5\%.

Finally, we test the effect on parameters of varying aspects of the low-$\ell$ 
TT treatment. These are discussed in Appendix B, and in summary we
find the same parameter results for the pixel-based likelihood code 
compared to the Gibbs code, when both use $\ell \le 32$. 
Changing the mask at low-$\ell$ to KQ80, or using the Gibbs code 
up to $\ell\le 51$, instead 
of $\ell \le 32$, has a negligible effect on parameters.

\subsection{Consistency of the $\Lambda$CDM model with other data sets}
\label{subsec:data}

While the \map\ data alone place strong constraints on cosmological 
parameters, there has been a wealth of results from 
other cosmological observations in the last few years. 
These observations can generally be 
used either to show consistency of the simple $\Lambda$CDM model 
parameters, or to constrain more complicated models. 
In this section we compare a broad range of current astronomical data 
to the \map\ $\Lambda$CDM model. A subset of the data is used to 
place combined constraints on extended cosmological models in 
\citet{komatsu/etal:prep}. For this subset, we describe the 
methodology used to compute the likelihood for each case, but do not 
report on the joint constraints in this paper.

\subsubsection{Small-scale CMB measurements}
\label{subsec:cmb}

\begin{figure*}[t]
  \epsscale{0.7}
  \plotone{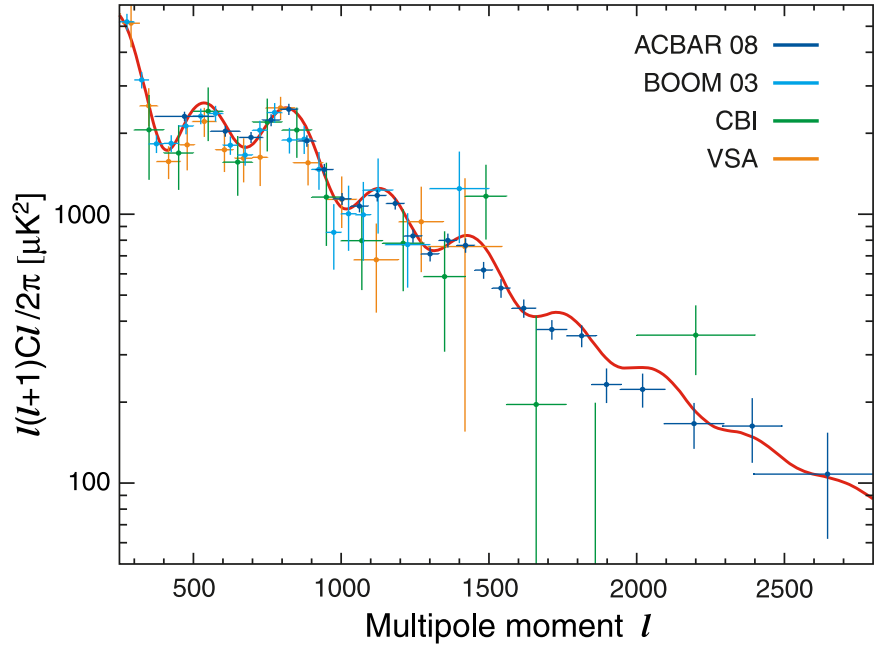}
  \caption{The best-fit temperature angular power spectrum from \map\ alone (red), 
is consistent 
with data from recent small-scale CMB experiments: ACBAR, CBI, VSA and 
BOOMERANG. 
    \label{fig:cmb_small} }
\end{figure*}

A number of recent CMB experiments have probed smaller angular scales 
than \map\ can reach and are therefore more sensitive to the higher order 
acoustic oscillations and the details of recombination (e.g., \citet{hu/sugiyama:1994,hu/sugiyama:1996,hu/white:1997}).
Since the three-year \map\ analysis, 
there have been new temperature results from 
the Arcminute Cosmology Bolometer Array Receiver (ACBAR), both in 2007 
\citep{kuo/etal:2007} and in 2008 \citep{reichardt/etal:prep}. They have 
measured the angular power spectrum at 145 GHz to 5' resolution, over $\sim$600 
deg$^2$. Their results are consistent with the model predicted by the 
\map\ five-year data, shown in Figure \ref{fig:cmb_small}, although 
ACBAR is calibrated using \map, so the data are not completely independent.

Figure \ref{fig:cmb_small} also shows data from the 
BOOMERANG, CBI and VSA experiments, which agree well with \map. 
There have also been new observations of the CMB polarization from two
ground-based experiments, QUaD, operating at 100 GHz and 150 GHz 
\citep{ade/etal:2007}, and CAPMAP, at 40 GHz and 100 GHz 
\citep{bischoff/etal:2008}. Their measurements of the 
EE power spectrum are shown in \citet{nolta/etal:prep}, together with 
detections already made since 2005 
\citep{leitch/etal:2005,sievers/etal:2007,barkats/etal:2005,montroy/etal:2006}, and are all consistent with the $\Lambda$CDM model parameters. 

In our combined analysis in \citet{komatsu/etal:prep} we use two different data
combinations. For the first we combine four data sets. This includes 
the 2007 ACBAR data \citep{kuo/etal:2007}, using 10 bandpowers in 
the range $900 \lt \ell \lt 2000$. The values and errors were obtained from 
the ACBAR web site.
We also include the three external CMB data sets used in 
\citet{spergel/etal:2003}: 
the Cosmic Background Imager 
(CBI; \citet{mason/etal:2003,sievers/etal:2003,pearson/etal:2003,readhead/etal:2004}), the Very Small Array 
(VSA; \citet{Dickinson/etal:2004}) and BOOMERANG 
\citep{ruhl/etal:2003,montroy/etal:2006,piacentini/etal:2006}. 
As in the three-year release we only use bandpowers that do not 
overlap with the 
signal-dominated \map\ data, due to non-trivial cross-correlations, so 
we use seven bandpowers for CBI (in the range $948<\ell<1739$), five for VSA 
($894<\ell<1407$) and seven for BOOMERANG ($924<\ell<1370$), 
using the lognormal form of the likelihood. Constraints are also found by  
combining \map\ with the 2008 ACBAR data, using 
16 bandpowers in the range $900<\ell<2000$. In this case the other CMB experiments
are not included.
We do not use additional polarization results for 
parameter constraints as they do not 
yet improve limits beyond \map\ alone.

\subsubsection{Baryon Acoustic Oscillations}
\label{subsec:bao}

\begin{figure*}[t]
  \epsscale{1.0}
  \plotone{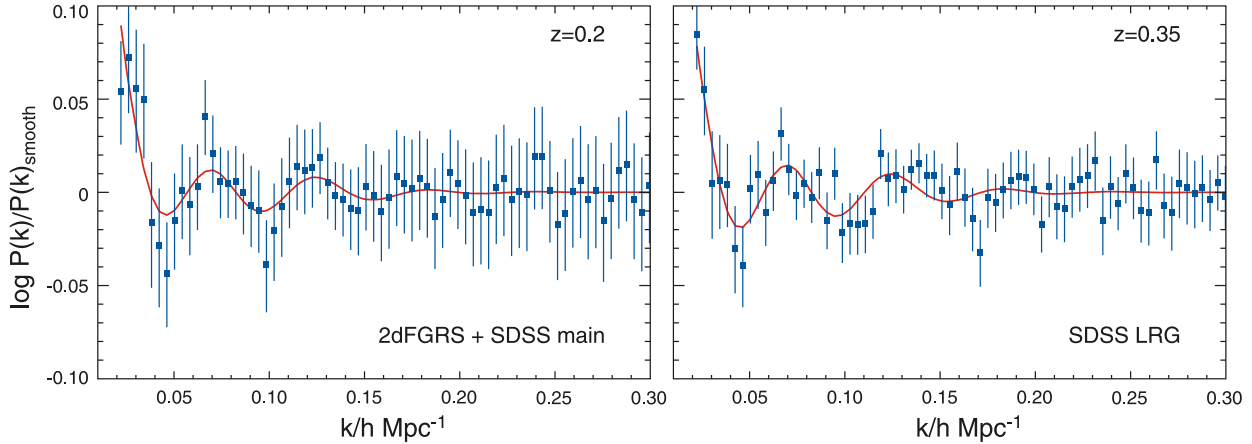}
  \caption{Baryon Acoustic Oscillations expected for the best-fit $\Lambda$CDM 
model (red lines), compared to BAO in galaxy power spectra 
calculated from (left) 
combined SDSS and 2dFGRS main galaxies, and (right) SDSS LRG galaxies, by 
\citet{percival/etal:2007c}. The observed and model power spectra have been 
divided by $P(k)_{\rm smooth}$, a smooth cubic spline fit described 
in \citet{percival/etal:2007c}. 
    \label{fig:bao_data} }
\end{figure*}

The acoustic peak in the galaxy correlation function is a prediction of the 
adiabatic cosmological model \citep{peebles/yu:1970,sunyaev/zeldovich:1970,bond/efstathiou:1984,hu/sugiyama:1996}. It was first detected using the 
SDSS luminous red galaxy (LRG) survey, using the brightest class of 
galaxies at mean redshift $z=0.35$ by \citet{eisenstein/etal:2005}. 
The peak was detected at $100h^{-1}$Mpc 
separation, providing a standard ruler to measure the ratio of 
distances to $z=0.35$ and the CMB at 
$z=1089$, and the absolute distance to $z=0.35$.
More recently \citet{percival/etal:2007a} have obtained 
a stronger detection from over half a million SDSS main galaxies and LRGs in 
the DR5 sample. They detect baryon acoustic oscillations (BAO) 
with over 99\% confidence.
A combined analysis was then undertaken of 
SDSS and 2dFGRS by \citet{percival/etal:2007c}. They find evidence for BAO in 
three catalogs: at mean redshift $z=0.2$ in the SDSS DR5 main 
galaxies plus the 2dFGRS galaxies, 
at $z=0.35$ in the SDSS LRGs, and in the combined catalog. Their data are shown 
in Figure \ref{fig:bao_data}, together with the \map\ best-fit model. 
The BAO are shown by dividing the observed and model power 
spectra by $P(k)_{\rm smooth}$, a smooth cubic spline fit described 
in \citet{percival/etal:2007c}. The observed power spectra are model-dependent, 
but were calculated 
using $\Omega_m=0.25$ and $h=0.72$, which agrees with our maximum-likelihood 
model.
  
The scale of the BAO is analyzed to estimate the 
geometrical distance measure at $z=0.2$ and $z=0.35$,
\be
D_V(z)=[(1+z)^2D_A^2cz/H(z)]^{1/3},
\ee
where $D_A$ is the angular diameter distance and $H(z)$ 
the Hubble parameter. They find $r_s/D_V(0.2)=0.1980\pm0.0058$ and 
$r_s/D_V(0.35)=0.1094\pm0.0033$. Here $r_s$ is 
the comoving sound horizon scale at recombination.
Our $\Lambda$CDM model, using the \map\ data alone, 
gives $r_s/D_V(0.2)=$ \ensuremath{0.1946\pm 0.0079}~and 
$r_s/D_V(0.35)=$ \ensuremath{0.1165\pm 0.0042}, 
showing the consistency  
between the CMB measurement at $z=1089$ and the late-time galaxy clustering.
However, while the $z=0.2$ measures agree to within 1$\sigma$, 
the $z=0.35$ measurements have mean values 
almost 2$\sigma$ apart.
The BAO constraints are tighter then the \map\ predictions, 
which shows that they can improve upon the \map\ parameter 
determinations, in particular on $\Omega_\Lambda$ and $\Omega_mh^2$.

In \citet{komatsu/etal:prep} the combined bounds from 
both surveys are used to constrain models as described 
in \citet{percival/etal:2007c}, adding a likelihood term given by 
$-2\ln L= X^T C^{-1}X$, with
\be
X^T= [r_s/D_V(0.2)-0.1980, r_s/D_V(0.35)-0.1094]\\
\ee
and $C^{11}=35059$, $C^{12}=-24031$, $C^{22}=108300$,
including the correlation between the two measurements.
\citet{komatsu/etal:prep} also consider constraints 
using the SDSS LRG limits derived by 
\citet{eisenstein/etal:2005}, using the combination 
\be
A(z)=D_V(z)\sqrt{\Omega_mH_0^2}/cz
\ee
for $z=0.35$ and computing a Gaussian likelihood
\be 
-2\ln L=(A-0.469(n_s/0.98)^{-0.35})^2/0.017^2.
\ee

\subsubsection{Galaxy power spectra}
\label{subsec:spectra}

We can compare the predicted fluctuations from the CMB 
to the shape of galaxy power spectra, in addition to the 
scale of acoustic oscillations (e.g., \citet{eisenstein/hu:1998}). 
The SDSS galaxy power spectum from DR3
\citep{tegmark/etal:2004} and the 2dFGRS spectrum \citep{cole/etal:2005} were
shown to be in good agreement with the \map\ three-year data, and used 
to place tighter constraints on  cosmological models 
\citep{spergel/etal:2007}, but there was 
some tension between the preferred values of the matter 
density ($\Omega_m=0.236\pm0.020$ for \map\ combined 
with 2dFGRS and $0.265\pm0.030$ with SDSS).
Recent studies used photometric redshifts to estimate the galaxy power 
spectrum of LRGs in the range $0.2 <z<0.6$  from the 
SDSS fourth data release (DR4), finding 
$\Omega_m = 0.30\pm0.03$ (for $h=70$, \citet{padmanabhan/etal:2007}) and 
$\Omega_m h = 0.195\pm0.023$ for $h = 0.75$ \citep{blake/etal:2007}. 

More precise measurements of the LRG power spectrum were obtained 
from redshift measurements: \citet{tegmark/etal:2006} used 
LRGs from SDSS DR4 in the range $0.01h/{\rm Mpc}<k<0.2h/{\rm Mpc}$
combined with the three-year \map\ data to place strong 
constraints on cosmological models.  
However, there is a disagreement between the matter density predicted using 
different minimum scales, if 
the non-linear modeling used in \citet{tegmark/etal:2006} is adopted. 
Using the three-year \map\ data combined with the LRG 
spectrum we find $\Omega_m=0.228\pm0.019$, using scales with 
$k<0.1h$~Mpc$^{-1}$, and $\Omega_m=0.248\pm0.018$ for $k<0.2h$~Mpc$^{-1}$. 
These constraints are obtained for the 6 parameter $\Lambda$CDM model, 
following the non-linear prescription in \citep{tegmark/etal:2006}. 
The predicted galaxy power spectrum $P_g(k)$ is constructed
from the `dewiggled' linear matter power spectrum $P_m(k)$ using 
$P_g(k) = b^2 P_m(k) (1+Qk^2)/(1+1.4k)$, and the parameters 
$b$ and $Q$ are marginalized over. The dewiggled spectrum is a weighted 
sum of the matter power spectrum at $z=0$ and a smooth power spectrum 
\citep{tegmark/etal:2006}. 
Without this dewiggling, we find $\Omega_m=0.246\pm0.018$ for 
$k<0.2h$~Mpc$^{-1}$, so its effect is small. 
We also explored an alternative form for the bias, motivated by third-order 
perturbation theory analysis, with 
$P_g(k)= b^2 P_m^{\rm nl}(k) + N$ (see e.g., \citet{seljak:2001,schulz/white:2006}).  Here $P_m^{\rm nl}$ is the 
non-linear matter power spectrum using the Halofit model 
\citep{smith/etal:2003} and $N$ is a constant accounting for non-linear 
evolution and scale-dependent bias.
Marginalizing over $b$ and $N$ we still find a discrepancy with scale, 
with $\Omega_m=0.230\pm0.021$ using scales with $k<0.1h$~Mpc$^{-1}$, and 
$\Omega_m=0.249\pm0.018$ for $k<0.2h$~Mpc$^{-1}$. Constraints using this
bias model are also considered in \citet{hamann/etal:prep}. 
These findings are consistent with results obtained from the 
DR5 main galaxy and LRG sample \citep{percival/etal:2007b}, who 
argue that this shows evidence for scale-dependent bias on 
large-scales, which could explain the observed differences in the early 
SDSS and 2dFGRS results. While this will likely be resolved with future 
modeling and observations, we choose not to use the galaxy power 
spectra to place 
joint constraints on the models reported in \citet{komatsu/etal:prep}.

\begin{figure*}[t]
  \epsscale{0.6}
  \plotone{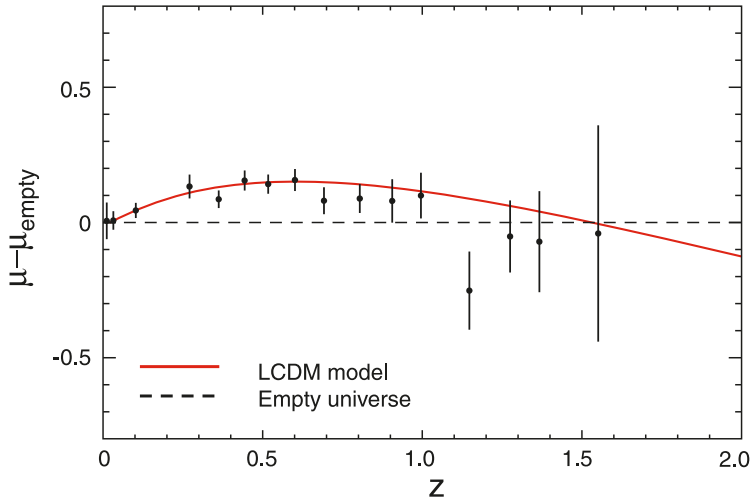}
  \caption{
The red line shows the luminosity distance relationship predicted for the best-fit WMAP-only model (right column in Table 2).  
The points show binned supernova observations from the Union compilation 
\citep{kowalski/etal:prep},
including high redshift SNIa from HST \citep{riess/etal:2007}, 
ESSENCE \citep{miknaitis/etal:2007}, and SNLS \citep{astier/etal:2006}.
The plot shows the deviation of the luminosity distances 
from the empty universe model. 
    \label{fig:sn_data} }
\end{figure*}

\subsubsection{Type Ia Supernovae}
\label{subsec:sn}

In the last decade Type Ia supernovae have become an important 
cosmological probe, and have provided the first 
direct evidence for the acceleration of the universe by
measuring the luminosity distance as a function of redshift. The 
observed dimness of high redshift supernovae ($z\sim0.5$) was first 
measured by 
\citet{riess/etal:1998,schmidt/etal:1998,perlmutter/etal:1999}, confirmed
with more recent measurements including \citet{nobili/etal:2005,krisciunas/etal:2005,clocchiatti/etal:2006,astier/etal:2006}, and extended to higher redshift
by \citet{riess/etal:2004} who found evidence for the earlier deceleration of 
the universe.
The sample of high redshift supernovae has grown by over 80 since the 
three-year \map\ analysis. Recent HST measurements of 21 new high 
redshift supernova by \citet{riess/etal:2007} include 13 at $z>1$, 
allowing the measurement of the Hubble expansion $H(z)$ at distinct 
epochs and strengthening the evidence for a period of deceleration 
followed by acceleration. The ESSENCE Supernova Survey 
has also recently reported results from 102 supernovae discovered from 
2002 to 2005 using the 4-m Blanco Telescope at the Cerro Tololo 
Inter-American Observatory \citep{miknaitis/etal:2007}, of which 60 
are used for cosmological analysis \citep{wood-vasey/etal:2007}. 
A combined cosmological analysis 
was performed of a subset of the complete supernova data
set by \citet{davis/etal:2007} using the MCLS2k2 light curve fitter 
\citep{jha/riess/kirshner:2007}.
More recently, a broader `Union' compilation of the currently observed 
SNIa, including a new nearby sample, was analysed by 
\citet{kowalski/etal:prep} using the SALT1 light curve fitter 
\citep{guy/etal:2005}.

We confirm in Figure \ref{fig:sn_data} that the recently observed 
supernovae are consistent with the $\Lambda$CDM model, which 
predicts the luminosity distance $\mu_{th}$ as a function of 
redshift and is compared to the Union combined data set 
\citep{kowalski/etal:prep}. In the cosmological analysis described in 
\citet{komatsu/etal:prep}, the Union data is used, consisting of 
307 supernovae that 
pass various selection cuts. These include supernovae observed using HST  
\citet{riess/etal:2007}, from the ESSENCE survey \citep{miknaitis/etal:2007}, 
and the Supernova Legacy Survey (SNLS, \citet{astier/etal:2006}).
For each supernova the luminosity distance predicted from theory is 
compared to the observed value. This is derived from measurements of the
apparent magnitude $m$ and the inferred absolute magnitude $M$, 
to estimate a luminosity distance $\mu_{obs}=5\log[d_L(z)/\rm Mpc]+25$. 
The likelihood is given by 
\be
-2\ln L=\sum_i[\mu_{obs,i}(z_i)-\mu_{th,i}(z_i,M_0)]^2/\sigma^2_{obs,i}
\ee
summed over all supernovae, 
where a single absolute magnitude is marginalized over 
\citep{lewis/bridle:2002} and $\sigma_{obs}$ is the observational 
error accounting for extinction, intrinsic redshift dispersion, 
K-correction and light curve stretch factors.

\subsubsection{Hubble constant measurements}
\label{subsec:hubble}

The \map\ estimated value of the Hubble constant, 
\ensuremath{H_0 = 71.9^{+ 2.6}_{- 2.7}}$~\rm km~\rm s^{-1}~\rm Mpc^{-1}$, 
assuming a flat geometry, 
is consistent with the HST Key Project measurement of 
$72\pm8~\rm km~\rm s^{-1}~\rm Mpc^{-1}$ \citep{freedman/etal:2001}. 
It also agrees within $1\sigma$ with measurements from gravitationally 
lensed systems \citep{koopmans/etal:2003}, 
SZ and X-ray observations \citep{bonamente/etal:2006}, Cepheid 
distances to nearby galaxies \citep{riess/etal:2005}, the distance to 
the Maser-host galaxy NGC4258 as a calibrator for 
the Cepheid distance scale \citep{macri/etal:2006}, and 
a new measure of the Tully-Fisher zero-point \citep{masters/etal:2006}, 
the latter two both giving $H_0=74\pm7~\rm km~\rm s^{-1}~\rm Mpc^{-1}$. 
Lower measures are favored by a compilation of Cepheid 
distance measurements for ten galaxies using HST by 
\citet{sandage/etal:2006} ($H_0=62\pm6~\rm km~\rm s^{-1}~\rm Mpc^{-1}$), and 
new measurements of an eclipsing binary in M33 which reduce the Key Project 
measurement to $H_0=61~\rm km~\rm s^{-1}~\rm Mpc^{-1}$
\citep{bonanos/etal:2006}. Higher measures are found 
using revised parallaxes for Cepheids \citep{van-leeuwen/etal:2007}, raising 
the Key Project value to 
$76\pm8~\rm km~\rm s^{-1}~\rm Mpc^{-1}$. 
In \citet{komatsu/etal:prep} the Hubble Constant measurements are included 
for a limited set of parameter constraints, using the 
\citet{freedman/etal:2001} measurement as a Gaussian prior on $H_0$.

\begin{table}[t] 
\begin{center}
\begin{tabular}{cccc}
\hline
\hline
Data & Parameter & Lensing limits & 5-year \map\ limits\\
\hline
CFHTLS Wide &  $\sigma_8(\Omega_m/0.25)^{0.64}$ & $0.785\pm0.043$ & $0.814\pm0.090$\\
100 Sq Deg & $\sigma_8(\Omega_m/0.24)^{0.59}$ & $0.84\pm0.07$ & $0.832\pm0.088$\\
COSMOS 2D & $\sigma_8(\Omega_m/0.3)^{0.48}$& $0.81\pm 0.17$ & $0.741\pm0.069$\\

COSMOS 3D & $\sigma_8(\Omega_m/0.3)^{0.44}$& $0.866^{+0.085}_{-0.068}$ & $0.745\pm0.067$\\
\hline
\end{tabular}
\caption{\small{Measurements of combinations of the matter density, $\Omega_m$, 
and amplitude of matter fluctuations, $\sigma_8$, from
weak lensing observations \citep{fu/etal:2008,benjamin/etal:2007,massey/etal:2007}, compared to \map.}
\label{table:lens}}

\end{center}
\end{table}

\subsubsection{Weak Lensing}
\label{subsec:lens}

Weak gravitational lensing is produced by the distortion of galaxy 
images by the mass distribution along the line of sight (see \citet{refregier:2003} for a review). 
There have been significant advances in its measurement in 
recent years, and in the understanding of systematic effects  (e.g., \citet{massey/etal:2007}), and intrinsic alignment 
effects \citep{hirata/etal:2007}, making it a 
valuable cosmological probe complementary to the CMB. 
Early results by the RCS \citep{hoekstra/etal:2002}, VIRMOS-DESCART 
\citep{vanwaerbeke/mellier/hoekstra:2005}, and the 
Canada-France-Hawaii Telescope Legacy Survey (CFHTLS,\citet{hoekstra/etal:2006}) lensing surveys favored higher 
amplitudes of mass fluctuations than preferred by \map.
However, new measurements of the two-point correlation functions 
from the third year CFHTLS Wide survey \citep{fu/etal:2008} favor a 
lower amplitude consistent with the \map\ measurements, as shown 
in Table \ref{table:lens}.
This is due to an improved estimate of the galaxy 
redshift distribution from CFHTLS-Deep \citep{ilbert/etal:2006}, compared to 
that obtained from photometric redshifts from the small Hubble Deep Field, 
which were dominated by systematic errors. 
Their measured signal agrees with results from the 100 Square 
Degree Survey \citep{benjamin/etal:2007}, a compilation of data 
from the earlier CFHTLS-Wide, RCS and VIRMOS-DESCAT surveys, together 
with the GABoDS survey 
\citep{hetterscheidt/etal:2007}, with average 
source redshift $z \sim 0.8$. Both these analyses rely on a 
two-dimensional measurement of the shear field. 
Cosmic shear has also been measured in two and three dimensions by the
HST COSMOS survey \citep{massey/etal:2007}, using redshift information to 
providing an improved measure of the mass fluctuation. 
Their measures are somewhat higher than the \map\ value, as shown in 
Table \ref{table:lens}, although not inconsistent.
Weak lensing is also produced by the distortion of the  
CMB by the intervening mass distribution 
\citep{zaldarriaga/seljak:1999,hu/okamoto:2002}, and 
can be probed by measuring 
the correlation of the lensed CMB with tracers of large 
scale structure. Two recent analyses have found 
evidence for the cross-correlated signal 
\citep{smith/zahn/dore:2007,hirata/etal:prep}, both 
consistent with the five-year \map\ $\Lambda$CDM model. They 
find a $3.4 \sigma$ detection of the correlation between
the three-year \map\ data and NVSS radio sources 
\citep{smith/zahn/dore:2007}, and a correlation at the 
$2.1 \sigma$ level of significance 
between \map\ and data from NVSS and SDSS LRGs and quasars 
\citep{hirata/etal:prep}.

\subsubsection{Integrated Sachs-Wolfe Effect}
Correlation between large-scale CMB temperature fluctuations and 
large-scale structure is expected in the $\Lambda$CDM model due to the 
change in gravitational potential as a function of time, and so provides a 
test for dark energy \citep{boughn/crittenden/turok:1998}. 
Evidence of a correlation was found in the 
first-year \map\ data (e.g., \citet{boughn/crittenden:2004,nolta/etal:2004}).
Two recent analyses combine recent large-scale structure data 
(2MASS, SDSS LRGs, SDSS quasars and NVSS radio sources) with the \map\ three-year data, finding a 
3.7$\sigma$ \citep{ho/etal:prep} and 4$\sigma$ \citep{giannantonio/etal:prep} 
detection of ISW at the expected level. 
Other recent studies using individual data sets find a correlation 
at the level expected with the 
SDSS DR4 galaxies \citep{cabre/etal:2006}, at high redshift with 
SDSS qusars \citep{giannantonio/etal:2006}, and with 
the NVSS radio galaxies \citep{pietrobon/etal:2006, mcewan/etal:2007}.

\subsubsection{Ly-$\alpha$ Forest}
The Ly$\alpha$ forest seen in quasar spectra probes the underlying matter 
distribution on small scales \citep{rauch:1998}.  However, the
relationship between absorption line structure and mass 
fluctuations must be fully understood to be used in a cosmological analysis.
The power spectrum of the Ly$\alpha$ forest has 
been used to constrain the shape and amplitude of the primordial 
power spectrum 
\citep{viel/weller/haehnelt:2004,mcdonald/etal:2005b,seljak/etal:2005,desjacques/nusser:2005}, and recent results combine the 
three-year \map\ data with the power spectrum
obtained from the LUQAS sample of VLT-UVES spectra 
\citep{viel/haehnelt/lewis:2006} and SDSS QSO spectra 
\citep{seljak/slosar/mcdonald:2006}. 
Both groups found results suggesting a higher value 
for $\sigma_8$ than consistent with \map.  
However, measurements by \citet{kim/etal:2007} of the 
probability distribution of the Ly$\alpha$ flux have 
been compared to simulations with different cosmological 
parameters and thermal histories \citep{bolton/etal:prep}. 
They imply that the temperature-density relation for the IGM may be close to 
isothermal or inverted, which would result in a smaller amplitude for 
the power spectrum than previously inferred, more in line with 
the five-year \map\ value of 
\ensuremath{\sigma_8 = 0.796\pm 0.036}.
Simulations in larger boxes by \citet{tytler/etal:prep} fail to match the
distribution of flux in observed spectra, providing further evidence of 
disagreement between simulation and observation.
Given these uncertainties, the Ly$\alpha$ forest data are not used 
for the main results presented in \citet{komatsu/etal:prep}. However, 
constraints on the 
running of the spectral index are discussed, using data described in 
\citet{seljak/slosar/mcdonald:2006}.
With more data and further analyses, the Ly$\alpha$ forest 
measurements can potentially place powerful constraints on 
the neutrino mass and a running spectral index.

\subsubsection{Big-Bang Nucleosynthesis}
\label{subsubsec:bbn}

\map\ measures the baryon abundance at decoupling, with
\ensuremath{\Omega_bh^2 = 0.02273\pm 0.00062}, 
giving a baryon to photon ratio of 
$\eta_{10}(\rm WMAP)=6.225\pm0.170$. Element abundances of deuterium, helium, 
and lithium also depend on the baryon abundance in the first 
few minutes after the Big Bang.
\citet{steigman:2007} reviews the current status of BBN measurements. 
Deuterium measurements provide the strongest test, and are consistent with 
\map, giving $\eta_{10}(D)=6.0\pm0.4$ based on new measurements by 
\citet{omeara/etal:2006}. 
The $^3$He abundance is more poorly constrained 
at $\eta_{10}(^3\rm He)=5.6^{+2.2}_{-1.4}$ from the measure 
of $y_3=1.1\pm0.2$ by
\citet{bania/rood/balser:2002}. The neutral lithium abundance, measured in 
low-metallicity stars, is two times smaller than the CMB 
prediction, $\eta(\rm Li)=$ from measures of the 
logarithmic abundance, $[\rm Li]_P = 12 +\log_{10}(\rm Li/\rm H)$, 
is the range $[\rm Li]_P \sim 2.1-2.4$ 
\citep{charbonnel/primas:2005,melendez/ramirez:2004,asplund/etal:proceedings}.
These measurements could be a signature of new early universe physics, 
e.g., \citep{coc/etal:2004,richard/michaud/richer:2005,jedamzik:2004}, with 
recent attempts to simultaneously fit both the $^7$Li and $^6$Li abundances by 
\citet{bird/koopmans/pospelov:2007,jedamzik:2008,cumberbatch/etal:2007,jedamzik:2008b,jittoh/etal:prep}. 
The discrepancy could also be due to systematics, destruction of 
lithium in an earlier 
generation of stars, or uncertainties in the stellar temparature scale 
\citep{fields/olive/vangioni-flam:2005,steigman:2006,asplund/etal:proceedings}.
A possible solution has been proposed using observations 
of stars in the globular cluster NGC 6397 \citep{korn/etal:2006,korn/etal:2007}. They find evidence that as the stars age and evolve towards hotter surface 
temperatures, the surface abundance of lithium in their atmospheres drops due
to atomic diffusion. They infer an initial lithium content of the stars 
$[\rm Li]_P=2.54\pm0.1$, giving $\eta_{10}(\rm Li)=5.4\pm0.6$ 
in good agreement with BBN predictions.
However, further observations are needed to determine whether this scenario 
explains the observed uniform depletion of primordial $^7$Li as a 
function of metallicity.
The measured abundance of $^4$He is also lower than predicted, with 
$\eta_{10}(^4\rm He)=2.7^{+1.2}_{-0.9}$ from a measure of 
$Y_P=0.240\pm0.006$, incorporating data from 
\citet{izotov/thuan:2004,olive/skillman:2004,gruenwald/steigman/viegas:2002} by
\citet{steigman:2007}. However, observations by 
\citet{peimbert/luridiana/peimbert:2007} and 
\citet{fukugita/kawasaki:2006} predict higher values more consistent with 
\map.

\subsubsection{Strong Lensing}

The number of strongly lensed quasars has the potential to 
probe cosmology, as a dark energy dominated universe 
predicts a large number of gravitational lenses 
\citep{turner:1990,fukugita/futamase/kasai:1990}. 
The CLASS radio band survey has a large statistical sample of 
radio lenses \citep{myers/etal:2003,koopmans/etal:2003,york/etal:2005}, 
yielding estimates for $\Omega_\Lambda \simeq  
0.72-0.78$ \citep{mitchell/etal:2005,chae:2007}.
\citet{oguri/etal:2008} have recently analyzed the large statistical lens  
sample from the Sloan Digital Sky Quasar Lens Search 
\citep{oguri/etal:2006}.  For a $w=-1$, flat
cosmology, they find $\Omega_\Lambda = 0.74_{-0.15}^{+0.11}({\rm  
stat.})_{-0.06}^{+0.13} ({\rm syst.})$.  These values are all consistent
with our best fit cosmology.
The abundance of giant arcs also has the potential to probe 
the underlying cosmology. However, although recent surveys have detected 
larger numbers of giant arcs 
\citep{gladders/etal:2003,sand/etal:2005,hennawi/etal:2008}
than argued to be consistent with 
$\Lambda$CDM \citep{li/etal:2006,broadhurst/barkana:prep}, 
numerical simulations \citep{meneghetti/etal:2007,hennawi/etal:2007,hilbert/etal:2007a,neto/etal:2007} find that the lens cross-sections are 
sensitive to the mass distribution in clusters 
as well as to the baryon physics 
\citep{wambsganss/ostriker/bode:prep,hilbert/etal:prep}.
These effects must be resolved in order to test $\Lambda$CDM consistency. 

\subsubsection{Galaxy clusters}
Clusters are easily detected and probe the high mass end of the mass
distribution, so probe the amplitude of density 
fluctuations and of large-scale structure. Cluster observations at 
optical wavelengths provided some of the first evidence for a low 
density universe with the current preferred cosmological parameters 
(see e.g., \citet{fan/bahcall/cen:1997}). Observers are now using a 
number of different techniques for identifying cluster samples: 
large optical samples, X-ray surveys, lensing surveys 
(see e.g., \citet{wittman/etal:2006}) and Sunyaev-Zeldovich surveys.
The ongoing challenge is to determine the
selection function and the relationship between astronomical observables 
and mass. This has recently seen significant 
progress in the optical \citep{lin/etal:2006,sheldon/etal:prep,reyes/etal:prep,rykoff/etal:prep} and the X-ray \citep{sheldon/etal:2001,reiprich/bohringer:2002,kravtsov/vikhlinin/nagai:2006,arnaud/pointecouteau/pratt:2007,hoekstra:2007}.
With large new optical cluster samples 
\citep{bahcall/etal:2003,hsieh/etal:2005,miller/etal:2005,koester/etal:2007} 
and X-ray samples from ROSAT, XMM-LSS and Chandra \citep{pierre/etal:2006,burenin/etal:2007,vikhlinin/etal:prep}, cosmological parameters can
be further tested, and most recent results for $\sigma_8$ are converging on 
values close to  the \map\ best fit value of 
\ensuremath{\sigma_8 = 0.796\pm 0.036}.
Constraints from the RCS survey \citep{gladders/etal:2007}
give  $\Omega_m = 0.30^{+0.12}_{-0.11}$ and $\sigma_8 = 0.70^{+0.27}_{-0.1}$.
\citet{rozo/etal:prep} argues for $\sigma_8 > 0.76$ 
(95\% confidence level) from SDSS BCG samples.
\citet{mantz/etal:2007} find $\Omega_m = 0.27^{+0.06}_{-0.05}$ and $\sigma_8 =0.77^{+0.07}_{-0.06}$
for a flat model based on the \citet{jenkins/etal:2001} mass function and
the \citet{reiprich/bohringer:2002} mass-luminosity calibration.
With a 30\% higher zero-point, \citet{rykoff/etal:prep} 
find that their data are best fit by $\sigma_8 = 0.85$
and $\Omega_m = 0.24$.  \citet{berge/etal:2007} report 
$\sigma_8 = 0.92_{-0.30}^{+0.26}$ for $\Omega_m = 0.24$ from 
their joint CFHTLS/XMM-LSS analysis.
In \citet{rines/diaferio/natarajan:2007} redshift data from SDSS DR4 is used 
to measure virial masses for a large sample of X-ray
clusters. For $\Omega_m=0.3$, they find $\sigma_8 = 0.84\pm0.03$.

\subsubsection{Galaxy peculiar velocities}

With deep large-scale structure surveys, 
cosmologists have now been able measure $\beta$, the amplitude 
of redshift space distortions as a function of redshift.  
Combined with a measurement of the bias, $b$, this yields a 
determination of the growth rate of structure 
$f \equiv d\ln G/d\ln a = \beta b$, where $G$ is the growth factor.
For Einstein gravity theories we expect 
$f \simeq \Omega^\gamma$ with $\gamma \simeq 6/11$ (see 
\citet{polarski/gannouji:prep} for a more accurate fitting function).
Analysis of redshift space distortions in the 2dF galaxy redshift 
survey \citep{peacock/etal:2001,verde/etal:2002,hawkins/etal:2003} 
find $\beta = 0.47 \pm 0.08$ at $z \simeq 0.1$, consistent
with the $\Lambda$CDM predictions for the best fit \map\ parameters.
The \citet{tegmark/etal:2006} analysis of the SDSS LRG sample 
finds $\beta = 0.31 \pm 0.04$ at $z = 0.35$.
The \citet{ross/etal:2007} analysis of the 2dF-SDSS LRG sample 
find $\beta = 0.45 \pm 0.05$ at z = 0.55.
The \citet{guzzo/etal:2008} analysis of  10,000 galaxies in the 
VIMOS-VLT Deep Survey finds that $\beta = 0.70 \pm 0.26$ at $z = 0.8$ and 
infer $d\ln G/d\ln a = 0.91 \pm 0.36$, consistent with the more 
rapid growth due to matter domination at this epoch expected in a $\Lambda$CDM 
model. The \citet{da-angela/etal:2008} analysis of a QSO sample 
finds $\beta = 0.60_{-0.11}^{+0.14}$ at $z =1.4$ and
use the clustering length to infer the bias.  Extrapolating 
back to $z = 0$, they find a matter density of $\Omega_m =  0.25^{+0.09}_{-0.07}$. A second approach is to use objects with well-determined 
distances, such as galaxies and supernovae, to look for 
deviations from the Hubble flow \citep{strauss/willick:1995,dekel:2000,zaroubi/etal:2001,Riess/press/kirshner:1995,haugbolle/etal:2007}. 
The \citet{park/park:2006}  analysis of the peculiar 
velocities of galaxies in the SFI sample find $\sigma_8 \Omega_m^{0.6} =
0.56 ^{+0.27}_{-0.21}$. \citet{gordon/land/slosar:2007} correlate
peculiar velocities of nearby supernova and find $\sigma_8 =0.79 \pm 0.22$.
These measurements provide an independent consistency check of the 
$\Lambda$CDM model (see \citet{nesseris/perivolaropoulos:2008} for a recent review).

\section{Extended cosmological models with \map\ }
\label{sec:ext_results}

The \map\ data place tight constraints on the simplest $\Lambda$CDM model 
parameters. In this section we describe to what extent \map\ data constrain 
extensions to the simple model, in terms of quantifying the 
primordial fluctuations and determining the composition of the universe beyond the 
standard components. \citet{komatsu/etal:prep} present constraints 
for \map\ combined with other data, and offer a more detailed cosmological 
interpretation of the limits.

\begin{figure*}[t]
  \epsscale{1.0}
  \plotone{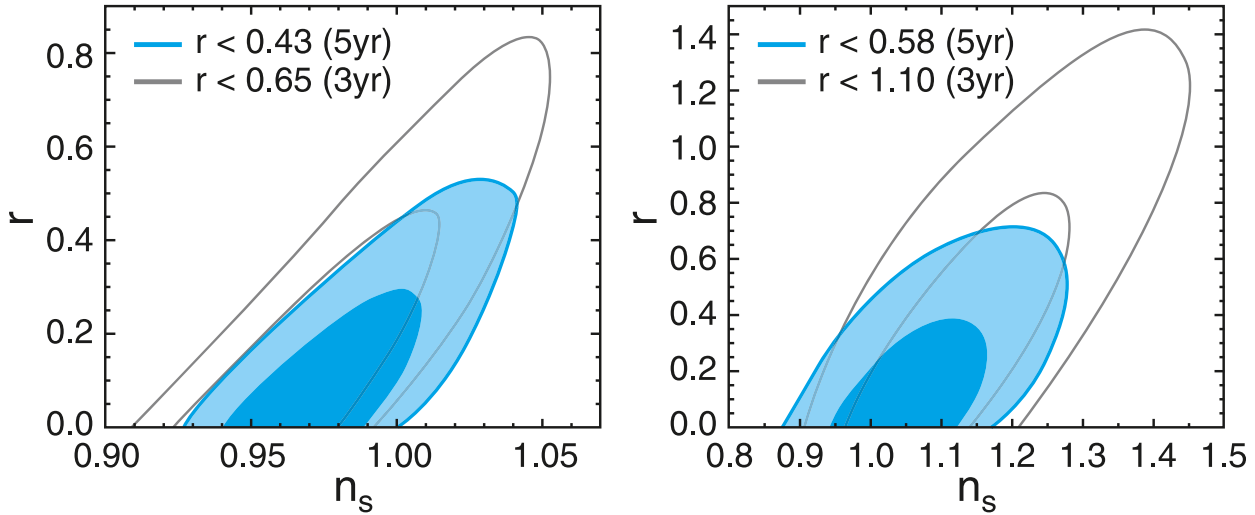}
  \caption{Two-dimensional marginalized constraints 
(68\% and 95\% confidence levels) on inflationary parameters $r$, 
the tensor-to-scalar ratio, and $n_s$, the spectral index of fluctuations, 
defined at $k_0=0.002/\rm Mpc$. One-dimensional 95\% upper limits on $r$ are given in the legend.
Left: The five-year \map\ data places 
stronger limits on $r$ (shown in blue) than three-year data (grey). 
This excludes some inflationary models 
including $\lambda \phi^4$ monomial inflaton models 
with $r\sim0.27$, $n_s\sim0.95$ for 60 e-folds of inflation. 
 Right: For models with a possible running spectral index, $r$ is now more
tightly constrained due to measurements of the third acoustic peak. Note: the 
two-dimensional 95\% limits correspond to $\Delta (2\ln L) \sim6$, 
so the curves intersect the $r=0$ line at the $\sim 2.5 \sigma$ limits of the 
marginalized $n_s$ distribution.
\label{fig:twod_tens} }

\end{figure*}

\begin{figure*}[t]
  \epsscale{0.6}
  \plotone{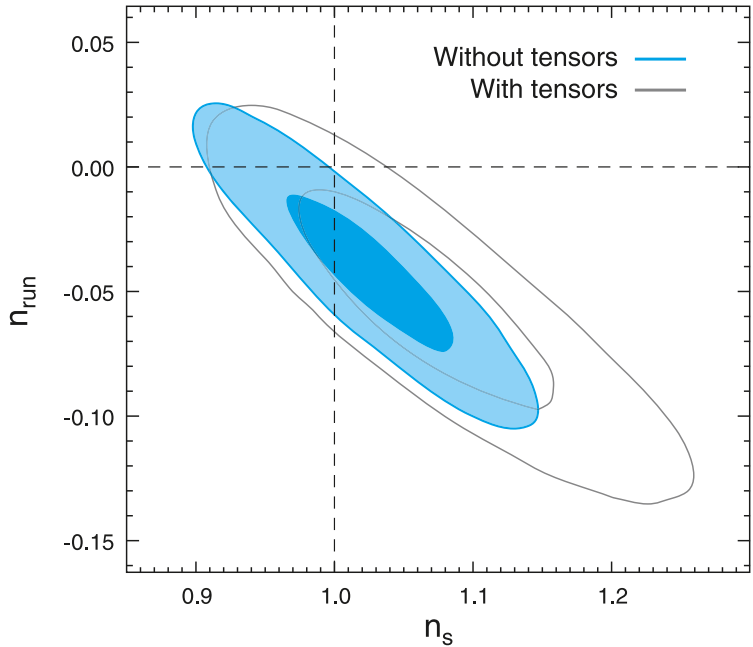}
  \caption{Two-dimensional marginalized limits for the spectral index, $n_s$, 
 defined at  $k_0=0.002/\rm Mpc$, and the running of the index $dn_s/d\ln k$ 
(marked $n_{\rm run}$). 
Models with no tensor contribution, and with a tensor contribution 
marginalized over, are shown. In both cases the models are consistent 
with a power-law spectral index, with $dn_s/d\ln k=0$, as expected from
the simplest inflationary models.
  \label{fig:twod_run} }
\end{figure*}

\subsection{Primordial perturbations}
\label{subsec:res_infl}

\subsubsection{Tensor fluctuations}

In the $\Lambda$CDM model, primordial scalar fluctuations are 
adiabatic and Gaussian, and 
can be described by a power law spectrum, 
\be
\Delta_{\cal R}^2(k) \propto \left(\frac{k}{k_0}\right)^{n_s-1}, 
\ee
producing CMB angular power spectra consistent 
with the data. Limits can also be placed on the amplitude of 
tensor fluctuations, or gravitational waves, that could have been generated
at very early times. They leave a distinctive large-scale 
signature in the polarized B-mode of the CMB (e.g., 
\citet{basko/polnarev:1980,bond/efstathiou:1984}), 
that provides a clean way to distinguish them from scalar fluctuations.
However, we have not yet reached sensitivities to strongly constrain 
this signal with the polarization data from \map. 
Instead we use the tensor contribution
to the temperature fluctuations at 
large scales to constrain the tensor-to-scalar ratio $r$.
We define $r=\Delta_h^2(k_0)/\Delta_{\cal R}^2(k_0)$, where $\Delta_h^2$ is 
the amplitude of primordial gravitational waves (see \citet{komatsu/etal:prep}), and choose a pivot scale $k_0=0.002/{\rm Mpc}$.

The \map\ data 
now constrain \ensuremath{r < 0.43\ \mbox{(95\% CL)}}. 
This is an improvement over the three-year limit of $r<0.65$ (95\% CL), 
and comes from 
the more accurate measurement of the second and third acoustic peaks. 
The dependence of the tensor 
amplitude on the spectral index is shown in Figure \ref{fig:twod_tens}, showing the
$n_s-r$ degeneracy \citep{spergel/etal:2007}: a larger contribution from 
tensors at large scales can be offset by an increased spectral index, 
and an overall decrease in the amplitude of fluctuations, shown in 
Table \ref{table:tens}. The degeneracy
is partially broken with a better measure of the TT spectrum.
There is a significant improvement in the limit on models whose scalar 
fluctuations can vary with scale, with a power spectrum with a `running' 
spectral index, 
\be
n_s(k)=n_s(k_0)+ \frac{1}{2}\frac{dn_s}{d\ln k} \ln \left(\frac{k}{k_0}\right).
\label{eqn:running}
\ee
The limit from \map\ is now \ensuremath{r < 0.58\ \mbox{(95\% CL)}}, 
about half the three-year value $r<1.1$ \citep{spergel/etal:2007}.

\begin{table} [t] 
\begin{center}
\begin{tabular}{cccc}
\hline
\hline
Parameter & Tensors & Running & Tensors+Running \\
\hline
$r$ & \ensuremath{< 0.43\ \mbox{(95\% CL)}} & &\ensuremath{< 0.58\ \mbox{(95\% CL)}} \\
$dn_s/d\ln k$ & &\ensuremath{-0.037\pm 0.028} & \ensuremath{-0.050\pm 0.034} \\
$n_s$ & \ensuremath{0.986\pm 0.022} & \ensuremath{1.031^{+ 0.054}_{- 0.055}} & \ensuremath{1.087^{+ 0.072}_{- 0.073}}\\
$\sigma_8$ & \ensuremath{0.777^{+ 0.040}_{- 0.041}} & \ensuremath{0.816\pm 0.036} & \ensuremath{0.800\pm 0.041}\\
\hline
\end{tabular}
\caption{\small{Selection of cosmological parameter constraints for 
extensions to the $\Lambda$CDM model including tensors and/or a running 
spectral index.}
\label{table:tens}}
\end{center}
\end{table}

What do these limits tell us about the early universe? 
For models that predict 
observable gravitational waves, it allows us to exclude more of 
the parameter space. The simplest inflationary models predict a 
nearly scale-invariant spectrum of gravitational waves 
\citep{grishchuk:1975,starobinsky:1979}.
In a simple classical scenario 
where inflation is driven by the potiential $V(\phi)$ of a 
slowly rolling scalar field, the predictions \citep{lyth/riotto:1999} are
\ba
r &\simeq & 4\alpha/N\\
1-n_s &\simeq & (\alpha+2)/2N
\ea
for $V(\phi) \propto \phi^\alpha$, where N is the number of 
$e$-folds of inflation between the time when horizon scale 
modes left the horizon and the end of inflation. For N=60, the 
$\lambda\phi^4$ model with $r\simeq 0.27$, $n_s\simeq 0.95$ is 
now excluded with more than 95\% confidence. An $m^2\phi^2$ model 
with $r\simeq 0.13$, $n_s\simeq 0.97$ is still consistent with the data. 
\citet{komatsu/etal:prep} discuss in some detail 
what these measurements, and constraints for combined data sets, 
imply for a large set of possible inflationary models and potentials.

With $r=0$ also fitting the data well, models that do not predict 
an observable level of gravitational waves, including multi-field 
inflationary models \citep{polarski/starobinsky:1995,garcia-bellido/wands:1996}, D-brane inflation 
\citep{baumann/mcallister:2007}, 
and ekpyrotic or cyclic scenarios 
\citep{khoury/etal:2001,boyle/steinhardt/turok:2004}, are not excluded if one fits for 
both tensors and scalars.

\subsubsection{Scale dependence of spectral index}

The running of the spectral index has been the subject of some debate 
in light of \map\ observations, with the three-year data giving 
limits of $dn_s/d\ln k= -0.055\pm0.030$, showing some 
preference for decreasing power on small scales \citep{spergel/etal:2007}. 
Combined with 
high-$\ell$ CBI and VSA CMB data, a 
negative running was preferred at $\sim 2\sigma$. 
A running index is not predicted by the 
simplest inflationary models (see e.g., \citet{kosowsky/turner:1995}), 
and the detection of a scale dependence
would have interesting consequences for early universe models. Deviations
from a power law index, and their consequences, have been considered by a 
number of groups in light of three-year data, including 
\citet{easther/peiris:2006,kinney/kolb/melchiorri:2006,shafieloo/souradeep:2007,verde/peiris:prep}, using various parameterizations.
In this analysis, and in \citet{komatsu/etal:prep} we consider only a running
index parameterized as in Eqn \ref{eqn:running}.
 
We show in Figure \ref{fig:twod_run} that with a better determination of the 
third acoustic peak, coupled with improved beam determination, the 
five-year \map\ data do not significantly 
prefer a scale-dependent index. The limit on the running is 
\ensuremath{dn_s/d\ln{k} = -0.037\pm 0.028} for models with no 
tensor contribution. 
The running is anti-correlated with the 
tensor amplitude \citep{spergel/etal:2003,spergel/etal:2007}, so the positive 
prior on the tensor amplitude leads to a more negative running preferred, 
\ensuremath{dn_s/d\ln{k} = -0.050\pm 0.034},
when a tensor contribution is marginalized over. Both limits are 
well within 2$\sigma$ of zero, showing no evidence of 
departure from a power law spectral index.

\subsubsection{Entropy perturbations}
\label{subsubsec:iso}

\begin{figure*}[tb]
  \epsscale{1.0}
  \plotone{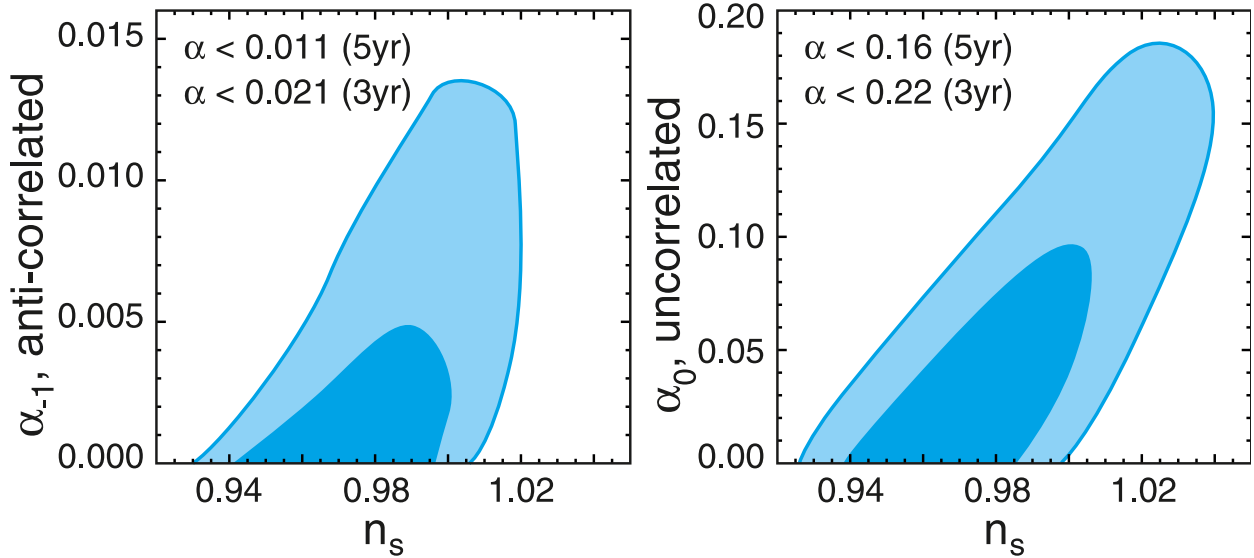}
  \caption{Marginalized two-dimensional limits (68\% and 95\%) on 
the amplitude of possible CDM entropy (or isocurvature) fluctuations. The 
one-dimensional 95\% upper limits are given in the legend. 
Left: Anti-correlated fluctuations are tightly constrained, 
placing limits on curvaton models. Right: Uncorrelated fluctuations, 
corresponding to axion models for dark matter, add less power to the CMB
spectrum than the anti-correlated case, for a given $\alpha$, so higher
values of $\alpha_0$ are allowed (than $\alpha_{-1}$), by the data. 
In both cases the amplitude is correlated with the spectral 
index of curvature fluctuations $n_s$, which compensates for 
the large scale power added by the CDM entropy fluctuations. 
  \label{fig:twod_iso} }
\end{figure*}

The simplest classical single-field inflation models predict 
solely adiabatic fluctuations, but entropy (or isocurvature) 
fluctuations are also predicted in a wide range of scenarios, 
including axions \citep{seckel/turner:1985,linde:1985}, multi-field inflation 
\citep{polarski/starobinsky:1994,garcia-bellido/wands:1996,linde/mukhanov:1997}, and decay of fields such as the 
curvaton \citep{lyth/wands:2002,moroi/takahashi:2001,bartolo/liddle:2002,lyth/ungarelli/wands:2003}. They may be correlated with the adiabatic fluctuations 
to some degree, depending on the model. Most physical scenarios generate 
only CDM or baryon entropy fluctuations 
\citep{bond/efstathiou:1987,peebles:1987}, with perturbation 
\be
{\cal S}_c = \frac{\delta \rho_c}{\rho_c} - \frac{3\delta\rho_\gamma}{4\rho_\gamma} , 
\ee
for CDM with density $\rho_c$. Neutrino modes are also possible 
\citep{bucher/moodley/turok:2000}. It has been known for some time that the CMB data cannot be fit by 
pure entropy fluctuations \citep{stompor/banday/gorski:1996,langlois/riazuelo:2000}, but a contribution may be allowed. 
Several groups have placed limits on a variety of models using 
the \map\ one-year and three-year data \citep{peiris/etal:2003,valiviita/muhonen:2003,bucher/etal:2004,beltran/etal:2004, dunkley/etal:2005b,kurki-suonio/muhonen/valiviita:2005, lewis:prep,bean/dunkley/pierpaoli:2006,trotta:2007,keskitalo/etal:2007}, finding no strong evidence for entropy fluctuations. 
Significant levels have been found to be consistent with the CMB data 
\citep{bucher/etal:2004, moodley/etal:2004,bean/dunkley/pierpaoli:2006}, but 
require correlated admixtures of CDM and neutrino isocurvature 
perturbations that are hard to motivate physically.

Here, and in \citet{komatsu/etal:prep}, we quantify the 
relative contributions to the angular power spectrum 
following \citet{beltran/etal:2004,bean/dunkley/pierpaoli:2006}, with
\be
C_\ell =  (1-\alpha)C_\ell^{\cal R} + \alpha C_\ell^{\cal S} +2\beta \sqrt{\alpha(1-\alpha)}C_\ell^{X},
\label{eqn_cl}
\ee
summing the spectra from curvature fluctuations $C_\ell^{\cal R}$, entropy 
fluctuations $C_\ell^{\cal S}$ with power spectrum
\be 
\Delta_S^2(k)\delta^3(\mb k- \mb k') = (k/2\pi)^3\langle{\cal S}_c(\mb k){\cal S}_c(\mb k')\rangle,
\ee
and a cross-correlation spectrum 
$C_\ell^{X}$ with power spectrum 
\be
\Delta_{X}^2 (k)\delta^3(\mb k- \mb k')= (k/2\pi)^3\langle-{\cal R}(\mb k) {\cal S}_c(\mb k')\rangle.
\ee 
This follows the definition of the curvature perturbation $\cal R$ 
in \citet{komatsu/etal:prep}, which 
gives large-scale temperature anisotropy $\Delta T/T=-{\cal R}/5$. An anti-correlated spectrum with $\beta=-1$ gives a positive $C_\ell^X$ on large scales. 

\begin{table} [t] 
\begin{center}
\begin{tabular}{ccc}
\hline
\hline
Parameter & $\beta=-1$  & $\beta=0$ \\
\hline
$\alpha_{-1}$ & \ensuremath{< 0.011\ \mbox{(95\% CL)}} &\\
$\alpha_{0}$ & &\ensuremath{< 0.16\ \mbox{(95\% CL)}}\\
$n_s$ &  \ensuremath{0.983\pm 0.017} & \ensuremath{0.987\pm 0.022}\\
$\sigma_8$ &  \ensuremath{0.778^{+ 0.039}_{- 0.038}} & \ensuremath{0.777\pm 0.038}\\
\hline
\end{tabular}
\caption{{\small Subset of cosmological parameter constraints 
for $\Lambda$CDM models 
with additional anti-correlated ($\beta=-1$) or uncorrelated ($\beta=0$) 
entropy fluctuations.}
\label{table:iso}}
\end{center}
\end{table}

Limits are found for $\alpha_{-1}$, corresponding to 
anti-correlated models with $\beta=-1$. This could correspond to a 
curvaton scenario in which inflation is driven by an inflaton field, but CDM 
perturbations are 
generated by the decay of a distinct curvaton field 
(see e.g., \citet{lyth/wands:2002}).   
In this case we make the assumption that the spectral index of the 
anti-uncorrelated entropy fluctuations is  equal to the adiabatic spectral 
index.
We do not find evidence for curvaton entropy pertubations, finding a
limit from 
\map\ of \ensuremath{\alpha_{-1} < 0.011\ \mbox{(95\% CL)}}, shown in 
Figure \ref{fig:twod_iso} and in Table \ref{table:iso}.
This is half the three-year limit, and places strong lower limits on 
the possible density of the curvaton at its decay, in 
this scenario, compared to the 
total energy density \citep{komatsu/etal:prep}. If the curvaton 
dominated at decay, the perturbations would be purely adiabatic.
Where has the improvement come from? The pure 
entropy spectrum and the cross-correlated spectrum both add large scale power, 
so a similar degeneracy is seen with the spectral index, and $\sigma_8$, as 
in the case where the tensor amplitude is varied. 
The entropy spectrum is also out of phase with the 
adiabatic spectrum, so the improved TE measurements combine with the 
third peak TT spectrum to tighten the limits. 

We also place limits on $\alpha_0$, corresponding to an uncorrelated 
model with $\beta=0$. In this case the entropy spectral index is set to be scale
invariant.
\citet{komatsu/etal:prep} describe how this corresponds to 
entropy perturbations created by axions, which would 
constitute some part of the dark matter budget.
The limit is 
\ensuremath{\alpha_{0} < 0.16\ \mbox{(95\% CL)}}, 
ten times higher than the anti-correlated amplitude, but still 
preferring pure adiabatic fluctuations. 
Without the large-scale power contribution from the 
anti-correlated spectrum, a much larger amplitude is permitted, but with 
the same degeneracy with the spectral index and $\sigma_8$. This has
implications for the maximum deviation from adiabaticity of 
axion dark matter and photons.
\citet{komatsu/etal:prep} provide a discussion of 
the theoretical implications of these limits, and those for combined data, 
for both models considered.

\subsection{Composition and geometry of the Universe}

\subsubsection{Number of relativistic species}

\begin{figure*}[t]
  \epsscale{1.0}
  \plotone{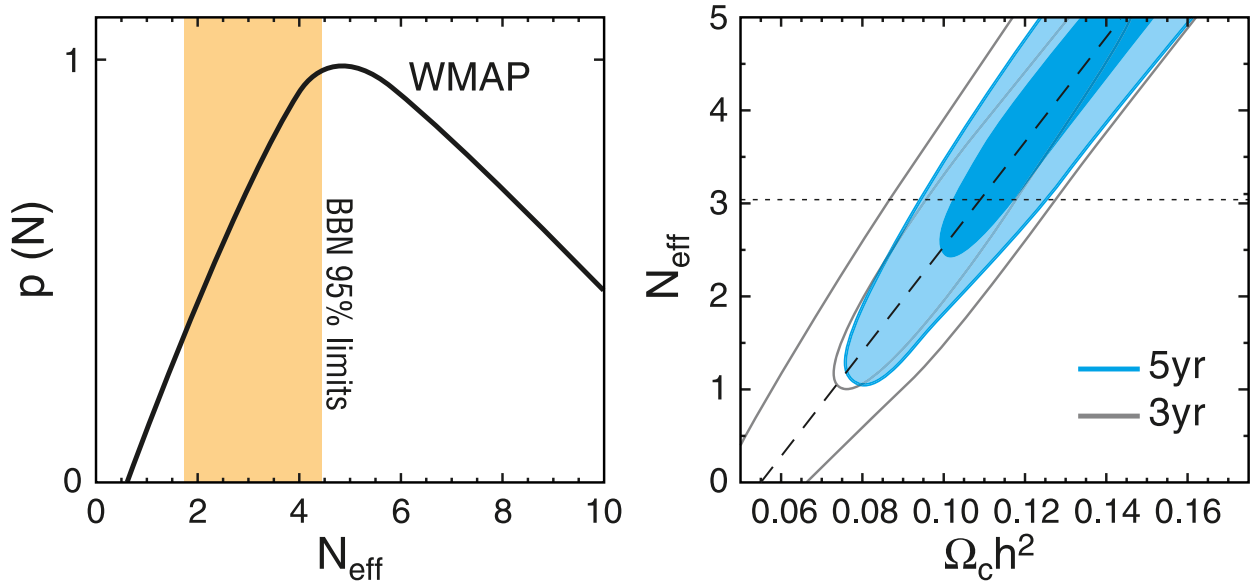}
    \caption{Evidence for a non-zero effective number of neutrino species, 
$N_{\rm eff}$. Left: The 
marginalized probability distribution gives 
\ensuremath{N_{\rm eff} > 2.3\ \mbox{(95\% CL)}} from \map\ alone. 
The best-fit $\Lambda$CDM model with $N_{\rm eff}=0$ is a poorer fit to 
the data than $N_{\rm eff}=3$, with $\Delta \chi^2=8.2$. Inferred 95\% limits from 
big bang nucleosynthesis (BBN) observations are highlighted.
Right: Joint two-dimensional distribution for $N_{\rm eff}$ and the 
CDM density, $\Omega_c h^2$, with five-year limits in blue, compared to three-year limits in grey. The degeneracy valley of constant $z_{eq}$ is shown dashed,
indicating that the CMB is now sensitive to the effect of neutrino anisotropic
stress, which breaks the degeneracy.  
  \label{fig:oned_neff} }
\end{figure*}

\begin{figure*}[t]
  \epsscale{0.6}
  \plotone{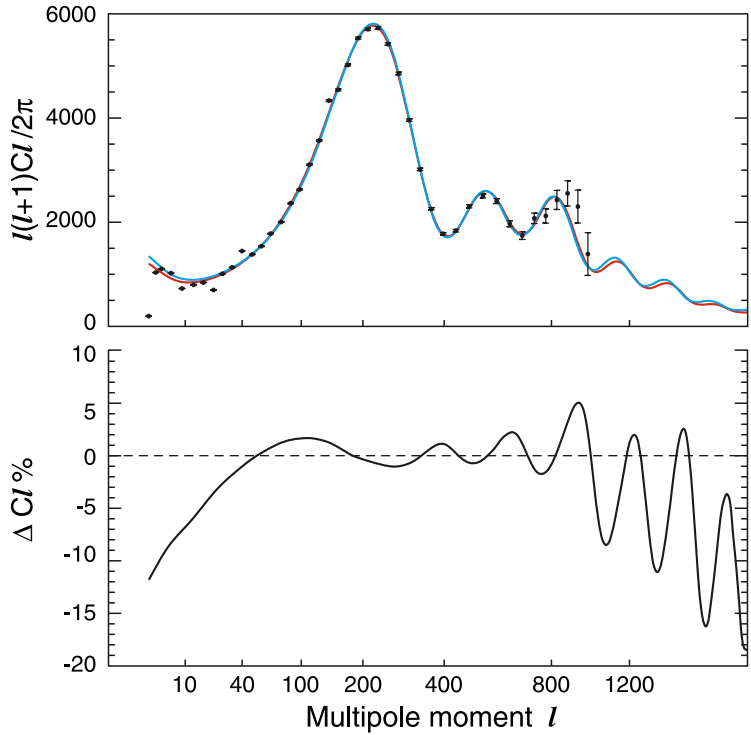}
   \caption{Comparison of the CMB angular power spectrum for the best-fit $\Lambda$CDM models with the standard $N_{\rm eff}=3.04$ neutrino species (red), 
and with $N_{\rm eff}=0$ species (blue). The lower panel shows the fractional
difference between the two spectra when $N_{\rm eff}$ is increased 
from $0$ to $3.04$. The $N_{\rm eff}=0$ model has a lower 
$\Omega_mh^2$ in order to fit the third peak,  and a lower spectral index, $n_s$, 
compared to the $N_{\rm eff}=3.04$ model.
  \label{fig:spec_neff} }
\end{figure*}

Neutrinos are expected to play an important role in the dynamics of the 
early universe.
For standard parameters, they contribute about 40\% of the energy 
density of the universe
during the radiation epoch and about 11\% of the energy density of the universe at
$z \sim 1100$ (very close to the energy density in baryons).
Because neutrinos contribute to the expansion of the universe and 
stream relativistically 
out of density fluctuations, they produce a significant imprint on the 
growth rate of structure and on the structure of the 
microwave background fluctuations (e.g., \citet{ma/bertschinger:1995}).  
The amplitude of
these effects depend upon $N_{\rm eff}$, the number of effective 
neutrino species. By `effective neutrinos species', 
we are counting any particle that is relativistic at
$z \sim 1000 - 3000$, couples very weakly to the baryon-electron-photon fluid,
and has very weak self-interactions.  Because we know neutrinos 
exist, we associate `neutrinos' with `light relativistic particle', 
but note that in the strictest sense we limit only 
light relativistic species, as the cosmological constraints are 
sensitive to the existence of {\it any} light species 
produced during the big bang or any
additional contribution to the energy density of 
the universe (e.g., primordial magnetic fields).

Measurements of the width of the Z provide very tight limits on the 
number of neutrino species:
$N_\nu = 2.984 \pm 0.008$ (Particle Data Book), consistent with the 
3 light neutrino species in the standard model.
Because of non-thermal effects due to the partial heating of 
neutrinos during the $e^\pm$ annihilations, and other small corrections, 
the effective number of
species is 3.0395 \citep{dicus/etal:1982,mangano/etal:2002}.  
Most analyses of the number of neutrino species with three-year \map\ data
\citep{spergel/etal:2007,ichikawa/kawasaki/takahashi:2007,mangano/etal:2007,hamann/etal:2007,debernardis/etal:2008} 
relied on combining CMB measurements with 
probes of the growth rate of structure. Since one
of the signatures of the number of neutrino species is a change in 
the growth rate of structure, there are 
degeneracies between the properties of the neutrinos and of the dark 
energy.
Neutrinos, however, leave a distinctive signature directly on the CMB power spectrum
(see \citet{bashinsky/seljak:2004} for detailed discussion): the neutrinos not 
only suppress the CMB peak heights, they also shift the acoustic 
peak positions.  While the effects that depend on shifts in the epoch 
of matter/radiation equality are degenerate with changes in the matter 
density, the effects of neutrino free-streaming are distinct.  Changes in 
the baryon/matter ratio and the baryon/photon ratio also have their own 
imprints on the Silk damping scale and on the acoustic scale.  With five
years of data, we are now able to see evidence of the effects of the 
neutrinos on the CMB power spectrum.

Figure 16 shows the limits on the number density of neutrinos and the 
density in dark matter. The degeneracy valley, shown in the right panel, 
corresponds to a constant 
ratio of matter density to radiation density, or equivalently a 
measurement of the expansion factor at matter radiation equality:
\ba
1 +z_{eq} &=& a_{eq}^{-1} =  \frac{\rho_c + \rho_{b}} {\rho_{\gamma} + \rho_{\nu}}\nonumber \\
	  &\simeq& 40500 \ \frac{\Omega_c h^2 + \Omega_b h^2}{1 + 0.23 N_{\rm eff}}.
\ea
With only 3 years of data and a lack of precision on the third peak 
position and height, \map\ was not able to make a clear detection of 
neutrinos (or relativistic species); 
however, the data did provide a $\sim 2 \sigma$ hint of the 
effects of neutrino anisotropic stresses \citep{melchiorri/serra:2006}.
Figure 16 shows that the five year data alone, we now 
constrain the number density of relativistic species:
\ensuremath{N_{\rm eff} > 2.3\ \mbox{(95\% CL)}}.  
By bounding $N_{\rm eff}<10$, and choosing a uniform 
prior on $N_{\rm eff}$, this level of significance depends  
somewhat on the prior.
We therefore test the significance of the constraint by 
comparing two $\Lambda$CDM models: one with $N_{\rm eff}=0$, 
and one with the standard $N_{\rm eff}=3.04$.
We find that the data prefer $N_{\rm eff}=3.04$. The best-fit model 
has $\Delta (-2\ln L) = 8.2$ less than the $N_{\rm eff}=0$ best-fit model, 
corresponding to evidence for relativistic species at $>99.5\%$ confidence.
The CMB power spectra corresponding to these two models, and 
their fractional difference, are shown in Figure 
\ref{fig:spec_neff}. The model with no neutrinos has a lower matter
density, $\Omega_mh^2$, in order to keep $z_{eq}$ fixed. 
The improvement in likelihood between the two models 
comes from both the low-$\ell$ and high-$\ell$ 
TT spectrum, with a small contribution from the TE spectrum. We also check that 
this evidence does not go away if we relax 
the assumption of a power-law spectral index, by
testing a model with a variable running, $dn_s/d\ln k \ne 0$.

\citet{komatsu/etal:prep} combine \map\ data with
other distance indicators (which constrain $\Omega_c h^2$) and finds a stronger
limit on the number density of neutrino species: 
\ensuremath{N_{\rm eff} = 4.4\pm 1.5}. 
These limits will continue 
to improve as CMB measurements of the higher peaks improve.
The CMB constraints on the number of relativistic species at redshift  
$\sim$1000-3000 complement constraints from big bang nucleosynthesis
and from particle accelerators.  Measurements of the abundance of 
Helium are sensitive to the expansion rate of the universe during its
first few minutes \citep{steigman/schramm/gunn:1977}. The agreement 
between the best fit value
from big bang nucleosynthesis, $N_{\rm eff} = 3.24 \pm 1.2$ 
(95\% confidence interval) (\citet{cyburt/etal:2005}, Particle Data Book 2007),
with the best fit CMB value is another consistency check 
for standard cosmology.  

\begin{table} [t] 
\begin{center}
\begin{tabular}{cc}
\hline
\hline
Parameter & Limits \\
\hline
$N_{\rm eff}$ & \ensuremath{> 2.3\ \mbox{(95\% CL)}}\\
$\sum m_\nu$ & \ensuremath{< 1.3\ \mbox{eV}\ \mbox{(95\% CL)}}\\
$Y_P$ & $\lt~0.45$ (95\% CL) \\
\hline
\end{tabular}
\caption{ \small{Constraints on neutrino properties and the primordial helium fraction.}
\label{table:nu}}
\end{center}
\end{table}

\subsubsection{Neutrino mass}

Cosmological data places limits on the mass of neutrinos. Atmospheric 
and solar neutrino experiments show that neutrinos are massive (see \citet{mohapatra/etal:prep}), 
and measure the difference between the square of their masses, $m_{\nu i}^2-m_{\nu j}^2$. 
Cosmological measurements constrain the sum of the masses $\sum m_\nu$ 
due to their
effect on the propagation of perturbations, on the clustering of matter, 
and on the expansion rate of the universe \citep{bond/szalay:1983,ma:1996,hu/eisenstein/tegmark:1998}. 
The mass has a large effect on 
the matter power spectrum, as massive neutrinos do not cluster as well as 
cold dark matter, leading to a suppression in power on small scales. 
Neutrinos also affect the CMB at earlier times: 
if the fraction of dark matter that is warm is raised, acoustic oscillations in the 
photon-baryon plasma are less strongly damped for modes that entered 
the horizon while the neutrinos were relativistic, 
raising the acoustic peak amplitudes. The radiation-like behavior at 
early times also changes the expansion rate, shifting the peak positions.

\begin{figure*}[t]
  \epsscale{1.0}
  \plotone{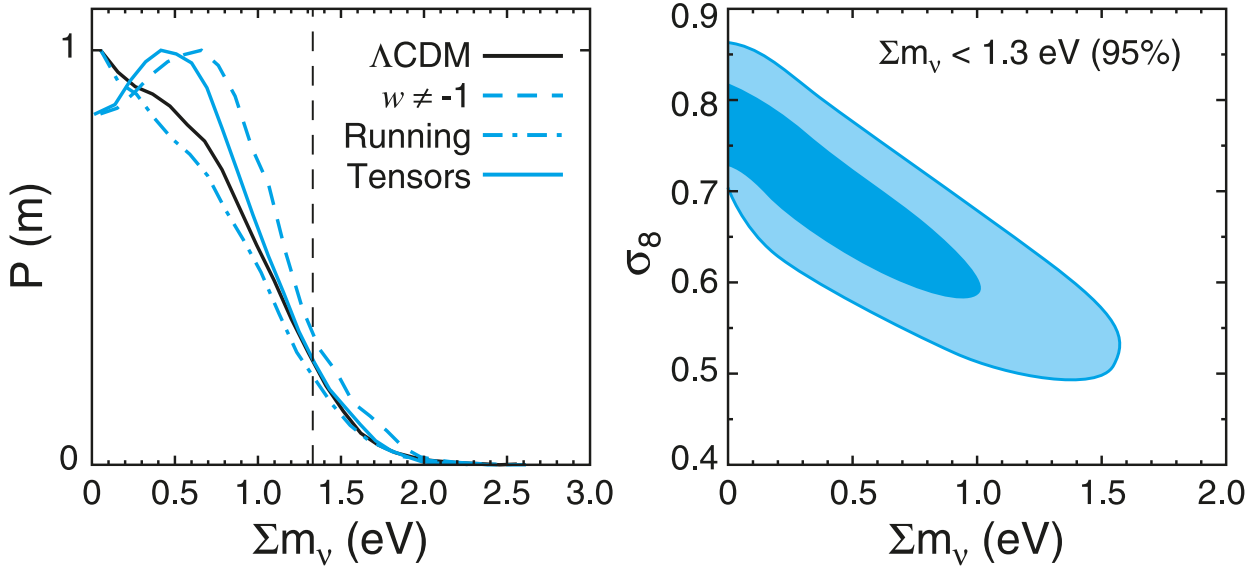}
   \caption{Limits on the sum of neutrino masses with the \map\ five-year data.
Left: The marginalized one-dimensional limit
from \map\ alone is \ensuremath{\sum m_\nu < 1.3\ \mbox{eV}\ \mbox{(95\% CL)}}. 
This is raised by $\lt 10\%$ with marginalization over a running spectral index, 
tensor fluctuations, or a dark energy equation of state $w$. 
Right: The neutrino mass is anti-correlated with $\sigma_8$, the 
amplitude of matter fluctuations. 
  \label{fig:oned_mnu} }
\end{figure*}

These effects are 
somewhat degenerate with other parameters, so
CMB data alone cannot limit the mass as well as when  
combined with other data. With the three-year \map\ data alone the limits 
were $\sum m_\nu <1.8$~eV \citep{spergel/etal:2007}, and $<0.66$~eV 
when combined with other data. Since the three-year 
\map\ analysis there have been
many studies of the constraints, as discussed in \citet{komatsu/etal:prep}. 

The five-year \map\ data now gives an upper limit on the total mass to be 
\ensuremath{\sum m_\nu < 1.3\ \mbox{eV}\ \mbox{(95\% CL)}}, shown in Table \ref{table:nu}.
We have checked that this upper limit is robust to the choice of 
cosmological models. The upper limit is raised  
by $< 10\%$ when we include tensor fluctuations, a 
running spectral index, or a constant $w\ne-1$ equation of state of 
dark energy, 
as shown in Figure \ref{fig:oned_mnu}. This dependence on additional parameters 
is consistent with earlier 
investigations 
by e.g., \citet{crotty/lesgourgues/pastor:2004,zunckel/ferreira:2007}. 
A larger neutrino 
mass raises the amplitude of the higher acoustic peaks, hence 
the observed degeneracy with $\sigma_8$ (Figure \ref{fig:oned_mnu}).
Stronger constraints come from combining the CMB data with probes of the 
expansion rate and clustering of matter at later times:
\citet{komatsu/etal:prep} find \ensuremath{\sum m_\nu < 0.61\ \mbox{(95\% CL)}} for \map\ combined with additional data.

\subsubsection{Primordial Helium Abundance}
In most cosmological analyses the primordial helium abundance is fixed to be 
$Y_P=0.24$, motivated by observations discussed in Sec \ref{subsubsec:bbn}. 
The effect of the abundance on the CMB spectrum is small, but provides an
independent cross-check of the BBN results, and probes for any 
difference between 
the helium abundance during the first few minutes, and after 300,000 years.
The abundance affects the CMB at small scales due to the recombination 
process. The number density of electrons before recombination depends on the 
helium fraction through $n_e=n_b(1-Y_P)$ where $n_b$ is the baryon 
number density. 
Changing the electron number density changes the mean 
free path of Compton scattering, which affects the Silk damping scale. 
A larger $Y_p$ increasingly damps the power on small scales, as shown in 
\citet{trotta/hansen:2004}. 

Constraints from the first-year \map\ data were presented in 
\citet{trotta/hansen:2004,huey/cyburt/wandelt:2004,ichikawa/takahashi:2006},
with 99\% upper limits of $Y_P<0.65$ inferred \citep{trotta/hansen:2004}.
A subsequent analysis of the three-year data gave $Y_P<0.61$ at 95\% 
confidence, tightened to 
$0.25\pm0.10$ with small-scale CMB data 
\citep{ichikawa/sekiguchi/takahashi:prep}.
We now find $Y_P<0.45$~(95\% CL)  with the five-year \map\ data. 
Higher values allowed by the thre-year data are disfavored with a 
better measure of the third acoustic peak height. With future 
small-scale CMB measurements, for example from the Planck satellite, 
constraints should significantly improve 
\citep{ichikawa/sekiguchi/takahashi:prep,hamann/lesgourgues/mangano:2008}.

\subsubsection{Curvature of the universe}
\begin{figure*}[tb]
  \epsscale{1.0}
  \plotone{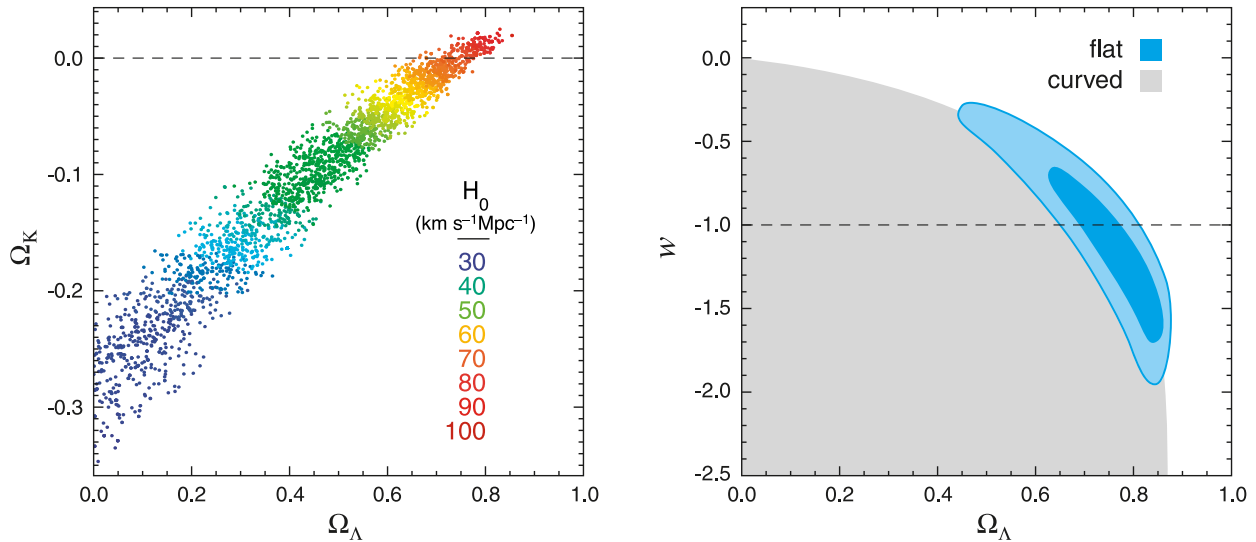}
  \caption{Left: The points show the 
set of non-flat models consistent with 
the \map\ data, colored by the Hubble constant values. 
\map\ measures 
the acoustic peak scale to high accuracy, but 
does not constrain the curvature, $\Omega_k$, by itself. However, the highly curved 
models have a low Hubble constant, inconsistent with observation. 
Right: Constraints on the dark energy equation of state, $w$, and the dark energy
density, $\Omega_\Lambda$, from \map\ alone. With a Hubble constant $H_0<100$, weak limits can be 
placed on $w$ in a flat universe, shown by the blue contours, but the 
dark energy density and equation of state are  
unconstrained (with the 95\% confidence level shaded grey) if 
the assumption of flatness is relaxed. Limits are significantly improved when \map\ 
is combined with additional data \citep{komatsu/etal:prep}.
  \label{fig:twod_geom} }
\end{figure*}

In combination with other data, \map\ observations place 
strong constraints on the geometry of the universe \citep{spergel/etal:2007}. 
The CMB measures with high accuracy the angular scale at which 
acoustic oscillations are imprinted at the last scattering surface, 
\ensuremath{\theta_{*} = 0.5952\pm 0.0017\ \mbox{\arcdeg}}. However, 
this alone does not provide a good measure of the geometry, as there is a 
degeneracy with the expansion rate of the universe since 
last scattering. This is 
shown in Figure \ref{fig:twod_geom}, indicating the degeneracy 
between the dark 
energy density $\Omega_\Lambda$ and the curvature $\Omega_k$. With \map\ alone 
the curvature is weakly constrained, with marginalized limits \ensuremath{\Omega_k = -0.099^{+ 0.085}_{- 0.100}}, and \ensuremath{\Omega_\Lambda < 0.76\ \mbox{(95\% CL)}},  assuming
a Hubble prior of $20 < H_0 < 100$ and $\Omega_\Lambda>0$. Without this
prior on the positivity of $\Omega_\Lambda$, limits on the curvature 
are weakened. 
The same degeneracy is seen, although slightly broadened, when the dark energy equation of state is allowed to vary. However, in both cases the Hubble constant 
decreases with increasingly negative curvature, taking values 
inconsistent with observation. This degeneracy can be used to 
constrain the curvature by combining observations \citep{jungman/etal:1996}. 
In the three-year \map\ analysis, \citet{spergel/etal:2007} showed 
that the degeneracy is truncated
with the addition of only one piece of additional cosmological data 
(Type Ia supernovae, or the HST measurement of the Hubble constant, or galaxy 
power spectra), tightly constraining any deviations from flatness. 
\citet{komatsu/etal:prep} draw similar conclusions with currently available 
data, and discuss the current limits on the spatial curvature 
from recent observations.

\subsubsection{Dark energy properties}
\label{subsec:dark}
The $\Lambda$CDM model requires a non-zero dark energy 
density \ensuremath{\Omega_\Lambda = 0.742\pm 0.030} to fit the data, which
is assumed to be in the form of a cosmological constant.
We do not have an explanation for this component of the universe. A 
natural explanation could be a vacuum energy density 
\citep{carroll/press/turner:1992}, but if so, we are faced with the 
fine-tuning problem to 
explain its observed value, 120 orders of magnitude smaller than expected 
from field theory arguments. 
Alternative explanations include quintessence 
\citep{peebles/ratra:1988,wetterich:1988,ferreira/joyce:1998} or 
modifications to gravity \citep{deffayet/dvali/gabadadze:2002}.
Testing the dark energy equation of state today, and 
as a function of 
cosmic time will help identify the possible explanation. 

The CMB by itself cannot place strong limits on the equation of state 
$w=p/\rho$, 
but by measuring the acoustic peak positions and heights, 
and constraining $\Omega_mh^2$ with the third peak, limits the range of models 
to a degeneracy between $\Omega_m$ and $w$, shown in 
Figure \ref{fig:twod_geom}. The dark energy in these models is allowed to 
cluster. With a prior on the Hubble constant $H_0<100$,
\map\ alone places weak limits \ensuremath{w = -1.06^{+ 0.41}_{- 0.42}}, with 
\ensuremath{\Omega_\Lambda = 0.73^{+ 0.10}_{- 0.11}}. If flatness is 
not assumed, the
\map\ data cannot constrain $w$ or $\Omega_\Lambda$ due to the geometric 
degeneracy, also shown in Figure \ref{fig:twod_geom}. 
However, the situation is significantly improved 
when \map\ is combined with astronomical data measuring the 
expansion rate and clustering of matter at late times. 
\citet{komatsu/etal:prep} discuss limits obtained from various data in 
combination with \map, and find $w$ constrained to be $-1$ to 
within 6\% for a flat universe and constant equation of state.

\section{Conclusions}
\label{sec:discuss}

The simple six parameter $\Lambda$CDM model continues to fit the \map\ data. 
With five years of observations, we have better measured both 
the temperature and polarization anisotropy of the CMB. 
This has allowed us to measure with smaller errors, compared to the three-year
analysis, the third acoustic peak in the temperature spectrum, and the low-$\ell$ 
polarization signal, leading to 
improved constraints on the cosmological parameters describing the contents 
of the universe, and the primordial fluctuations that seeded structure. 
The observations continue to be well fit by the predictions of the 
simplest inflationary models, with a scale-invariant spectrum of fluctuations 
disfavored. Consistency with the TE cross-correlation spectrum, now 
measured with better accuracy, provides 
additional confidence in this simple model.

We have detected the optical depth to reionization with high 
significance. This measurement implies that reionization of 
the universe likely took place gradually, as it constrains a 
sudden reionization to be earlier than consistent with other 
observations. With more data, it will become possible to use the 
polarization data to better quantify the ionization history.
Given the improvement in this measurement, and with a view to interpreting 
future large-scale polarization measurements, we develop an 
alternative way to remove Galactic foregrounds from low resolution 
polarization maps, which includes marginalization over uncertainties in the 
Galactic signal. We find consistent results using this method and the
standard template-cleaning method.

Considering a range of extended models, we continue 
to find that the standard $\Lambda$CDM model is consistently preferred 
by the data. The improved measurement of the third peak now 
requires the existence of light relativistic species, assumed to 
be neutrinos, at high confidence. 
The standard scenario 
has three neutrino species, but the three-year \map\ data could not rule 
out models with none. 
The $\Lambda$CDM model also continues to succeed in 
fitting a substantial array of other observations. Certain tensions between other
observations and those of \map, such as the amplitude of 
matter fluctuations measured by weak lensing surveys and using the 
Ly-$\alpha$ forest, and the primordial lithium abundance, have either 
been resolved with improved understanding of systematics, or show some 
promise of being explained by recent observations. 
With further \map\ observations 
we will better 
probe both the universe at a range of epochs, measuring fluctuation 
characteristics to probe the initial inflationary process, or other 
non-inflationary scenario, improving measurements of the composition of the 
universe at the recombination era, and characterizing 
the reionization process in the universe.

\acknowledgments

The \map\ mission is made possible by the support of the Science 
Mission Directorate Office at
NASA Headquarters.  This research was additionally 
supported by NASA grants NNG05GE76G,
NNX07AL75G S01, LTSA03-000-0090, ATPNNG04GK55G, and ADP03-0000-092.  
EK acknowledges support from an Alfred P. Sloan Research Fellowship. 
We thank Antony Lewis for discussion about lensing in CAMB, 
Will Percival for discussion and provision of BAO data, 
Catherine Heymans, Jonathan Benjamin, and Richard Massey for 
discussion of weak lensing data, Michael Wood-Vasey for 
discussion of ESSENCE supernova data, 
Eric Aubourg for discussion of SNLS supernova data, 
Gary Steigman for discussion of BBN constraints, 
Bruce Draine and Todd Thompson for discussion of dust and synchrotron 
emission.
This research has made 
use of NASA's Astrophysics Data System Bibliographic Services.  
We acknowledge use of the HEALPix, CAMB, CMBFAST, and CosmoMC packages.

\appendix
\label{pol_appendix}

\section{Low-$\ell$ TT likelihood cross checks}

\begin{figure}[t]
\begin{center}
  \epsscale{0.6}
  \plotone{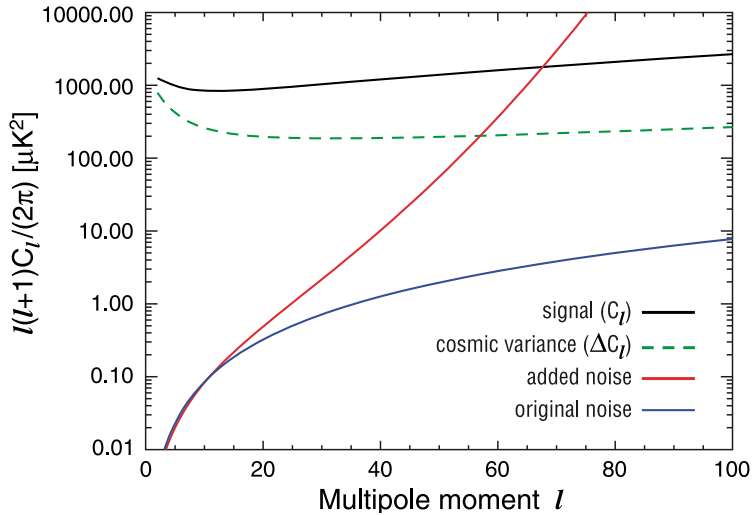}
\end{center}
\caption{The angular power spectra of signal and noise components in the
smoothed ILC map used for Gibbs sampling the low-$\ell$ temperature spectrum.  
Uncorrelated noise, at 2 $\mu K$ per pixel (red), is added to the smoothed 
ILC map to speed up the
sampling, and then is assumed to be the only noise present. This assumption is inaccurate at low
$\ell$, as it ignores the true noise (blue), but the error is negligible since it is significantly lower than cosmic variance (green).
\label{fig:gibbs_noise}}
\end{figure}

\begin{table}[b]
\begin{center}
\begin{tabular}{lccc}
\hline
resolution parameter & 4 & 5 \\
smoothing FWHM &  $9.1831^\circ$ & $5.0^\circ$ \\
$\sigma_{\rm noise}/{\rm pixel}$ [$\mu$K] & 1.0 & 2.0 \\
$\ell_{\rm max}$ sampled & 32 & 51 \\
$\ell_{\rm max}$ conditioned & 48 & 96 \\
\hline
\end{tabular}
\caption{{\small Parameters used for sampling the low-$\ell$ TT likelihood at two different resolutions.}
\label{res_table}}
\end{center}
\end{table}

\begin{figure*}[t]
  \epsscale{0.6}
  \plotone{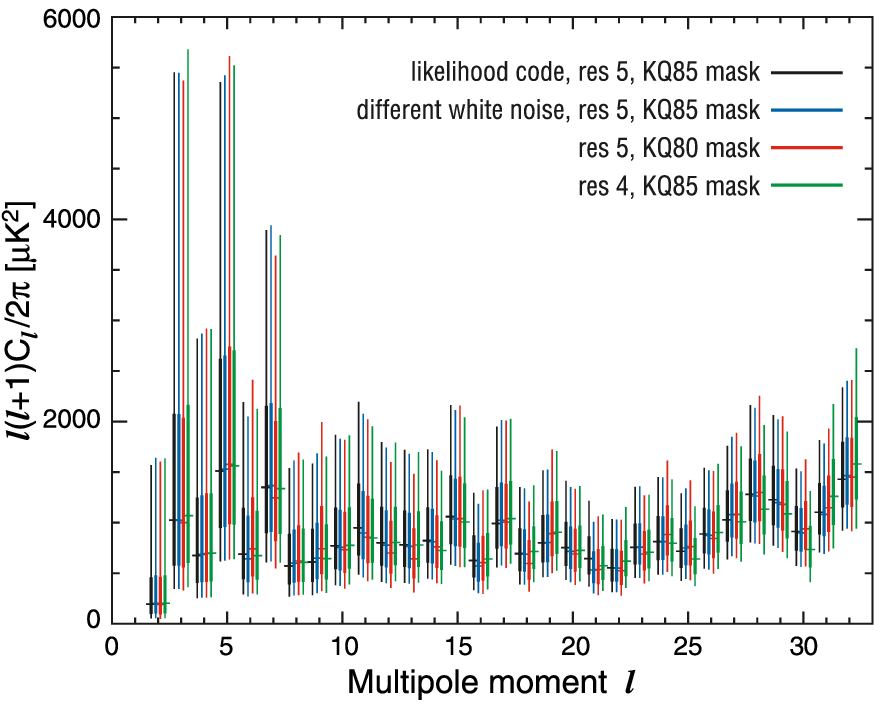}
  \caption{The level of variation in the low-ell TT Gibbs
likelihood that arises from different input parameters. The spectrum 
estimated from the standard input (black) is compared to results obtained 
using a different white noise realization (blue), 
using a larger mask (KQ80,red), and degrading to $N_{\rm side}=16$ (green) rather 
then $32$.  The
likelihood does fluctuate with these changes, but has a negligible effect 
on cosmological parameters. The values shown at each multipole 
correspond to maximum likelihood values obtained by fixing the spectrum of 
other multipoles at fiducial values. The error bars show where 
the likelihood is at 50\% and 5\% of its peak value.
\label{fig:gibbs_compare} }
\end{figure*}

We have switched from the pixel likelihood to a Blackwell-Rao
(BR) estimator at low ell, so we perform several tests to verify that the
new likelihood 1) is consistent with the pixel likelihood, 2)
is insensitive to several choices of input data, and 3) is properly
converged. Many of our checks involve comparing likelihoods.  Because the
likelihood is a function on a high dimensional space, we do not
check it everywhere.  Instead we choose some fiducial spectrum and
compute slices through the likelihood.  We compare these conditional
likelihoods, typically in the form of maximum likelihood values and
full-width-half-max error bars between different low-$\ell$ likelihood
methods.  We also compare $\Lambda$CDM cosmological parameters obtained using 
the different inputs.

The first comparison is between the pixel-based code and the Gibbs code.
Figure \ref{fig:pix_spec_check} showed a comparison of the 
$C_\ell$ likelihoods, with 
no significant differences, even though they use different resolutions.
The cosmological parameters 
are unchanged for the two codes. 
The next comparison is between likelihoods from Gibbs sampling 
at two different resolutions, $N_{\rm side} = 16$ and $32$.
 Table \ref{res_table} lists 
details of runs at these two 
different resolutions, including the standard deviation of 
Gaussian white noise added to the smoothed maps. The power spectrum of the 
white noise for the $N_{\rm side}=32$ case, is compared to the input noise and 
cosmic variance in Figure \ref{fig:gibbs_noise}. 
Power spectra are sampled up to $\ell_{\rm max}$ (sampled), 
and the sampled CMB skies are constructed with
the sampled power spectra and by conditioning on a constant fiducial
spectrum out to a higher value of $\ell_{\rm max}$ (conditioned).
In Figure \ref{fig:gibbs_compare} 
the estimated spectra are 
compared, using the fiducial KQ85 mask. There are some small 
differences, but the spectra are consistent.

We also show in Figure \ref{fig:gibbs_compare} 
that the spectra obtained are almost identical for 
two different realizations for the added uncorrelated white noise,
and that using the larger 
KQ80 mask, compared to the fiducial KQ85, has a small effect on the 
spectra and cosmological parameters, consistent with noise.
Finally we compare results for different $\ell_{\rm max}$ used in the 
BR estimator. The standard set-up 
uses $\ell_{\rm max}=32$; using 
the low-$\ell$ likelihood up to $\ell=51$ has almost no effect on cosmological 
parameters.

\section{Parameter estimation details}

\subsection{Sampling method}
\label{subsec:sample}

In the primary pipeline, only one chain is run for each model, rather 
than the typical four or eight parallel chains used in most MCMC cosmological 
analyses. This is possible as our spectral convergence test only 
requires a single chain. The starting points are picked as 
good-fitting points from previous analyses 
(e.g., a good-fitting value from the three-year \map\ analysis), 
or previous test chains, 
which means that no initial burn-in need be removed. Starting afresh with the
five-year data, one can always use a point lying in the \map\ three-year 
limits as a starting point. This would not be the case for entirely new 
distributions, or much improved data, in which case we run a short 
initial exploratory chain to find the high likelihood region. 

The covariance matrix from which trial steps are drawn, 
is chosen using a best guess for the covariance $C$ 
of the distribution being sampled.
For a multi-variate Gaussian with covariance $\mb C$, 
the optimal trial covariance matrix, if all parameters are sampled 
simultaneously, is 
${\mb C}_T=(\sigma _T/\sigma _0)^2{\mb C}$, 
where 
\be
\frac{\sigma_T}{\sigma_0}\approx \frac{2.4}{\sqrt{D}}
\ee 
for D-dimensional distributions 
\citep{gelman/roberts/gilks:1996,hanson/cunningham:1998,dunkley/etal:2005}. 
This relation also holds for somewhat non-Gaussian distributions.
At each step we draw a vector $\mb G$ of D Gaussian unit variance zero mean 
random variates, and compute a trial set of 
parameters $\mb x_T$, starting from 
the current position $\mb x_i$, where
\be
\mb x_T=\mb x_i+\sqrt \mb C_T \mb G.
\ee
Using an appropriate covariance matrix can speed up the sampling time by a 
factor of hundreds. In practice we have a good 
idea of the covariance from previous cosmological analyses, so this 
is always used as a starting point. With significantly different data, or 
with new parameters, a best guess is made.
The chains are then run for a few thousand steps, and then updated if the
matrix is inadequate. This is determined by the acceptance rate 
of the chain, which should be $\sim 15-25\%$ and by the chain efficiency,
using the spectral test described in the next section. 
A second or third update may be required for models such as those with
curved geometries with variable dark energy equation of state.
Once a good covariance matrix is found, the chains are 
run for typically 20,000 steps, and then tested for convergence. 
Convergence for $\Lambda$CDM chains typically 
takes place after only $\sim$6000 iterations, although at this point the 
1D and 2D distributions are noisy. In practice we then run all the 
chains for longer than the convergence limit,
typically for 100,000 iterations, in order to get well sampled 
histograms. The chains are not thinned before analysis.

\subsection{Spectral convergence test}
\label{subsec:conv}

A chain has `converged' when its statistical properties 
reflect those of the underlying distribution with sufficient 
accuracy. To determine this stopping point, we 
use the spectral convergence test described in \citet{dunkley/etal:2005} 
in our main parameter pipeline.
The power spectrum of each parameter of the chain is 
used as a diagnostic, to check whether the chain has (1) sampled the 
distribution in such a way that it is unbiased by correlations, and (2) 
sampled enough points that statistics can be estimated with sufficient 
accuracy. The Gelman \& Rubin test \citep{gelman/rubin:1992}, 
commonly used in cosmology, 
can sometimes fail to test the first point, producing a false positive.

To estimate the power spectrum $P(k)$ from a chain of length $N$ we 
construct ${\hat P}_j=\vert a_{j}\vert ^2$, where $j=2\pi k/N$, 
by taking the discrete Fourier Transform of the chain of values for 
each parameter, $x$,
\ba
a_j=\frac{1}{\sqrt{N}}\sum _{n=0}^{N-1} x(n)~\exp \bigl[ i2\pi (jn/N)\Bigr] 
\ea
where $-(N/2-1)<j<N/2$.
$x(n)$ is the value at each iteration $n$, 
so chains stored in weighted format are converted to unweighted 
arrays for analysis.
Since the Metropolis algorithm produces chains which are correlated on 
small scales, the power spectrum tends to a 
white noise spectrum on large scales, and turns 
over to a spectrum with suppressed power at large $k$, with the turnover 
position reflecting the inverse correlation 
length. In \citet{dunkley/etal:2005} it is shown that the spectrum 
can be fit by the following template:
\be{
P(k)=P_0\frac{(k^*/k)^\alpha}{(k^*/k)^\alpha+1}},
\label{template:e}
\ee
with $P_0$ giving the amplitude of the white
noise spectrum in the $k\to 0$ limit. $k^*$
indicates the position of the turnover to a different power law
behavior, characterized by $\alpha ,$ at
large $k.$  This model is shown in Figure 3 of \citet{dunkley/etal:2005}, 
fitting the noisy spectrum of a parameter from a chain. The model 
fits the noise-averaged spectrum of a real chain obtained from Monte-Carlo
simulations, also shown in \citet{dunkley/etal:2005}.

To fit the parameters $\ln P_0, k^*$ and $\alpha$ to ${\hat P}_j$ 
using least squares for a finite chain, we have
\ba
\ln {\hat P}_j = \ln [P_0]+\ln \left[
                   \frac{(Nk^*/2\pi j)^\alpha}{1+(Nk^*/2\pi j)^\alpha }
                    \right] -\gamma + r_j,
\ea
where $\gamma= 0.577216$ is the Euler-Mascheroni constant, and 
$r_j$ are random measurement errors. 
The parameters are fit over the range of Fourier modes 
$1\le j\le 10j^*$,  for a spectrum with $j^*=k^*(N/2\pi)$, so we
iterate twice to converge on the $j^*$ limit.

For convergence, the largest scales probed
must be in the white noise regime $P(k) \sim k^0$,
defined by the requirement $j^* > 20$ for each parameter. 
This insures that the 
correlated points are not biasing the distribution and indicates that the 
chain is drawing points throughout the full region of high probability.
To test sufficient accuracy, we require the convergence ratio  
$r=\sigma_{\bar{x}}^2/ \sigma_{0}^2$ to be less than 1\% for each parameter, 
where $\sigma^2_{\bar x}$ is the variance in the sample mean, and 
$\sigma_{0}^2$ is the variance of the sampled parameter. 
The Gelman \& Rubin test incorporates a 
similar ratio: their $R$ statistic roughly translates to $R \sim 1+r$, but 
the quantity is calculated using multiple parallel chains.
It is shown in \citet{dunkley/etal:2005} that $r$ can be 
estimated using a single chain, since estimating the sample mean variance 
of a long chain, with zero mean, is equivalent to estimating $P(k)$ at $k=0$:
\ba
\sigma^2_{\bar{x}}=\left< {{\bar x}}^2\right> \approx \frac{1}{N}\cdot P(k=0).
\ea
In practice we rescale each parameter to have zero mean and unit variance
before computing its power spectrum. Then we estimate the value of $P_0$ 
for each parameter, to compute $r= P_0/N$. We require $r<0.01$, but in practice 
obtain much smaller values, typically with $r<0.001$.

\end{document}